\documentclass[iop]{emulateapj}

\newcommand{\kms}{\,km\,s$^{-1}$}

\newcommand{\um}{\,$\mu$m}

\newcommand{\Sco}{$S_\nu({\rm CO})\Delta{ v}$}
\newcommand{\Scont}{$S_\nu (350)$}

\slugcomment{Accepted for publication in the Astrophysical Journal}

\begin{document}

\title{Blind detections of CO $J = 1-0$ in 11 H-ATLAS galaxies at
 $z=$\,2.1--3.5 with the GBT/Zpectrometer}

\author{A.I.~Harris\altaffilmark{1}, 
  A.J.~Baker\altaffilmark{2},
  D.T.~Frayer\altaffilmark{3}, 
  Ian~Smail\altaffilmark{4},  
  A.M.~Swinbank\altaffilmark{4},
  D.A.~Riechers\altaffilmark{5},
  P.P.~van~der~Werf\altaffilmark{6},
  R.~Auld\altaffilmark{7}, 
  M.~Baes\altaffilmark{8},  
  R.S.~Bussmann\altaffilmark{9},
  S.~Buttiglione\altaffilmark{10},
  A.~Cava\altaffilmark{11}, 
  D.L.~Clements\altaffilmark{12},
  A.~Cooray\altaffilmark{13}, 
  H.~Dannerbauer\altaffilmark{14, 15},
  A.~Dariush\altaffilmark{12, 7},
  G.~De~Zotti\altaffilmark{10}
  L.~Dunne\altaffilmark{16},
  S.~Dye\altaffilmark{17},
  S.~Eales\altaffilmark{7},
  J.~Fritz\altaffilmark{8},
  J.~Gonz\'{a}lez-Nuevo\altaffilmark{18},
  R.~Hopwood\altaffilmark{12},
  E.~Ibar\altaffilmark{19},
  R.J.~Ivison\altaffilmark{19, 20},
  M.J.~Jarvis\altaffilmark{21, 22},
  S.~Maddox\altaffilmark{16},
  M.~Negrello\altaffilmark{19, 10}, 
  E.~Rigby\altaffilmark{17}, 
  D.J.B.~Smith\altaffilmark{17},
  P.~Temi\altaffilmark{23}, 
  and J.~Wardlow\altaffilmark{13}} 

\altaffiltext{1}{Department of Astronomy, University of Maryland,
  College Park, MD 20742, USA; \emph{harris@astro.umd.edu}}

\altaffiltext{2}{Department of Physics and Astronomy, Rutgers, the
  State University of New Jersey, Piscataway, NJ 08854-8019, USA;
  \emph{ajbaker@physics.rutgers.edu}} 

\altaffiltext{3}{National Radio Astronomy Observatory, P.O.\ Box 2,
  Green Bank, WV 24944, USA; \emph{dfrayer@nrao.edu}}

\altaffiltext{4}{Institute for Computational Cosmology, 
  Department of Physics, Durham University, South Road, Durham DH1
  3LE, UK; \emph{ian.smail@durham.ac.uk}}  % smail, swinbank

\altaffiltext{5}{Astronomy Department, California Institute of
  Technology, MC 249-17, 1200 East California Boulevard, Pasadena, CA
  91125, USA}  % riechers

\altaffiltext{6}{Leiden Observatory, Leiden University, P.O. Box 9513, 
  NL-2300 RA Leiden, The Netherlands} % vanderwerf

\altaffiltext{7}{School of Physics and Astronomy, Cardiff University, 
  5 The Parade, Cardiff CF24 3AA, UK} 
  % auld, pascale, dariush, eales, pohlen

\altaffiltext{8}{Sterrenkundig Observatorium, Universiteit Gent,
  Krijgslaan 281 S9, B-9000 Gent, Belgium} % fritz, baes

\altaffiltext{9}{Harvard-Smithsonian CfA, 60 Garden Street, MS 78,
  Cambridge, MA 02138, USA} % bussmann

\altaffiltext{10}{INAF-Osservatorio Astronomico di Padova, Vicolo
  Osservatorio I-35122 Padova, Italy} % buttiglione, dezotti, negrello

\altaffiltext{11}{Departamento de Astrof\'{\i}sica, Facultad de
  CC. F\'{\i}sicas, Universidad Complutense de Madrid, E-28040 Madrid,
  Spain} % cava

\altaffiltext{12}{Physics Department, Imperial College London, South
  Kensington Campus, SW7 2AZ, UK} % dariush, clements, hopwood

\altaffiltext{13}{Department of Physics and Astronomy, University of
  California, Irvine, CA 92697, USA} % cooray, wardlow

\altaffiltext{14}{Universit\"{a}t Wien, Institut f\"ur Astronomie
  T\"{u}rkenschanzstra{\ss}e 17, 1180 Wien, Austria} % dannerbauer
\altaffiltext{15}{Laboratoire AIM-Paris-Saclay,
  CEA/DSM-CNRS-Universit\'{e} Paris Diderot, Irfu/SAp, CEA-Saclay, 
  91191 Gif-sur-Yvette Cedex, France} % dannerbauer

\altaffiltext{16}{Department of Physics and Astronomy, University of
  Canterbury, Private Bag 4800, Christchurch 8140, New Zealand} % dunne, maddox 
                                
\altaffiltext{17}{School of Physics and Astronomy, University of
  Nottingham, Nottingham, NG7 2RD, UK} % rigby, djbsmith, dye

\altaffiltext{18}{Instituto de Física de Cantabria, CSIC-UC, Av.\ de
  Los Castros s/n, Santander, 39005, Spain} %gonzalez

\altaffiltext{19}{UK Astronomy Technology Centre, Royal Observatory,
  Blackford Hill, Edinburgh EH9 3HJ, UK} % ibar, ivison

\altaffiltext{20}{Institute for Astronomy, University of Edinburgh, 
 Royal Observatory, Blackford Hill, Edinburgh EH9 3HJ, UK} % ivison

\altaffiltext{21}{Centre for Astrophysics Research, Science \&
  Technology Research Institute, University of Hertfordshire,
  Hatfield, Herts AL10 9AB; Physics Department} % jarvis
\altaffiltext{22}{University of the
  Western Cape, Cape Town 7535, South Africa} % jarvis

\altaffiltext{23}{Astrophysics Branch, NASA Ames Research Center, Mail
  Stop 245-6, Moffett Field, CA 94035, USA} % temi

\begin{abstract} 
  \noindent We report measurements of the carbon monoxide ground state
  rotational transition ($^{12}$C$^{16}$O $J = 1-0$) with the
  Zpectrometer ultra-wideband spectrometer on the 100\,m diameter
  Green Bank Telescope.  The sample comprises 11 galaxies with
  redshifts between $z = 2.1$ and 3.5 from a total sample of 24
  targets identified by {\em Herschel\/}-ATLAS photometric colors from
  the SPIRE instrument.  Nine of the CO measurements are new redshift
  determinations, substantially adding to the number of detections of
  galaxies with rest-frame peak submillimeter emission near 100\um.
  The CO detections confirm the existence of massive gas reservoirs
  within these luminous dusty star-forming galaxies (DSFGs).  The CO
  redshift distribution of the 350\um-selected galaxies is strikingly
  similar to the optical redshifts of 850\um-selected submillimeter
  galaxies (SMGs) in $2.1 \le z \le 3.5$.  Spectroscopic redshifts
  break a temperature-redshift degeneracy; optically thin dust models
  fit to the far-infrared photometry indicate characteristic dust
  temperatures near 34\,K for most of the galaxies we detect in CO.
  Detections of two warmer galaxies and statistically significant
  nondetections hint at warmer or molecule-poor DSFGs with redshifts
  difficult determine from from {\em Herschel}-SPIRE photometric
  colors alone.  Many of the galaxies identified by H-ATLAS photometry
  are expected to be amplified by foreground gravitational lenses.
  Analysis of CO linewidths and luminosities provides a method for
  finding approximate gravitational lens magnifications $\mu$ from
  spectroscopic data alone, yielding $\mu \sim 3$--20. Corrected for
  magnification, most galaxy luminosities are consistent with an
  ultra-luminous infrared galaxy (ULIRG) classification, but three are
  candidate hyper-LIRGs with luminosities greater than $10^{13} \;
  L_\odot$.
\end{abstract}


\section{Introduction}\label{sec:intro}

Observations of the far-IR/submillimeter background with {\it COBE}
demonstrated that a substantial fraction of the universe's star
formation took place behind a veil of dust \citep{puget96,fixsen98}.
Because these were integrated measurements, however, they could not
identify which populations of dust-obscured galaxies contained this
vigorous star formation.  A breakthrough in resolving the background
came in the late 1990s, with imaging by the Submillimeter Common-User
Bolometer Array (SCUBA) on the James Clerk Maxwell Telescope (JCMT).
Its initial surveys at 850\um\ \citep{smail97,barger98,hughes98}
detected a new population of bright ($> 5$\,mJy) galaxies.  Named
after the wavelengths where they are most visible, these submillimeter
galaxies (SMGs) are systems with apparently vast ($\gtrsim
10^{13}\,L_\odot$) bolometric luminosities but with such high obscurations
that their optical counterparts are faint or absent \citep[see][and
references therein]{blain02}.  SMGs brighter than SCUBA's confusion
limits could not account for all of the 850\um\ background, but
clearly made a substantial contribution to it.

Over the last fifteen years, much of the effort to understand the
origins of the far-IR/submillimeter background has focused on bright
SMGs selected from 850--1200\um\ surveys.  SMGs are sometimes treated
as representatives of a more general category of high-redshift
galaxies, dusty star-forming galaxies (DSFGs) whose luminosities are
dominated by obscured star formation.  A major initial hurdle was
verifying that bright SMGs actually do lie at high redshifts.  While
two early SCUBA detections had optical redshifts
\citep{ivison98,ivison00}, quickly confirmed with CO spectroscopy
\citep{frayer98,frayer99}, progress in measuring redshifts of other
SMGs foundered due to their very high obscurations.  Only after
\citet{chapman03} took advantage of radio continuum mapping to
determine precise positions for blind optical spectroscopy did it
become possible to obtain CO detections in large numbers
\citep{neri03,greve05,tacconi06}, and to derive an SMG redshift
distribution peaking around $z\sim 2$--2.5 \citep{chapman05}.  Heroic
efforts to explore the high-$z$ tail that radio pre-selection misses
when radio counterparts fall below typical survey flux limits have
identified a handful of sources at $z > 4$
\citep[e.g.,][]{capak08,coppin09, coppin10co,riechers10, capak11}, but detailed
analysis limits the possible significance of this tail in the $\sim
1$\,mm population \citep{ivison05, wardlow11}.  We now know that
bright SMGs have large stellar \citep{smail04,hainline10}, molecular
gas \citep{greve05,tacconi08}, and dynamical \citep{genzel03} masses,
that many of them are mergers \citep{conselice03}, and that their
large luminosities are powered principally by star formation
\citep{alexander03,alexander05,pope08}.  Explaining the observed
properties of bright SMGs, their evolutionary states, and their
relationships to populations of galaxies selected at other wavelengths
is a current major challenge for galaxy evolution models
\citep[e.g.,][]{baugh05,swinbank08,dave10,somerville11}.

In parallel with the growth in our understanding of bright SMGs, it is
also becoming clear that current samples give an incomplete picture of
the full variety of DSFGs.  First, 850\um\ sources fainter than
SCUBA's nominal confusion limit of about 2~mJy, although accessible
via gravitational lensing \citep[e.g.,][]{smail02,kneib05,knudsen08},
must at some point start to resemble optically selected galaxies more
than heavily obscured SMGs \citep[e.g.,][]{genzel10,daddi10}.  Second,
even among bright DSFGs, 850\um-bright SMGs have distinct selection
biases.  DSFG samples selected at longer wavelengths appear to have
cooler dust temperatures and higher median redshifts
\citep{dannerbauer04,valiante07,wardlow11,lindner11}.  DSFG samples
selected at shorter wavelengths, conversely, include populations with
warmer dust that 850\um-selected samples can miss,
\citep{blain04,chapman04}, and that tend to have both lower redshifts
and more bolometrically significant active galactic nuclei
\citep[AGNs; e.g.,][]{houck05,weedman06,yan07}.  Third, existing
samples of SMGs with spectroscopic redshifts often suffer from biases
associated with the steps used to determine those redshifts.  For
example, precise localization of a submillimeter source usually relies
on radio continuum mapping, and while the deepest VLA imaging yields
counterparts for a high fraction of DSFGs \citep{lindner11}, more
typical VLA maps tend to deliver counterparts for only 60--70\% of
SMGs.  Subsequent optical spectroscopy based on these positions fails
to yield redshifts for a modest fraction of candidates
\citep{chapman05}, and even when apparently successful, attempts to
obtain CO detections of the gas reservoirs associated with these
massive starbursts can sometimes fail to yield conclusive
confirmation, raising questions over either the identification or
redshift \citep{greve05}.  Mid-IR spectroscopy can avoid some of these
difficulties \citep[e.g.,][]{valiante07,menendezdelmestre07,
pope08,menendezdelmestre09,coppin10midir} but suffers from its own
problems for sources with power-law spectra or confusion from multiple
sources within large beams or slits.  Finally, many of the seminal
studies of SMGs mapped relatively small areas on the sky.
Notwithstanding the large line-of-sight interval probed by 850\um\
selection, which can partly compensate for a small area, SMGs are such
rare sources (mergers caught at special moments, with the most
luminous caught at the most special of moments) that cosmic variance
remains a concern for the derived redshift distributions.

New instruments capable of producing deep images of large regions of
the sky, and of conducting efficient spectral surveys over wide
bandwidths, have accelerated the discovery of high-redshift DSFGs with
a broader range of physical properties than could be probed by
previous efforts.  Survey areas at $\lambda \sim 1$\,mm have increased
substantially
\citep[e.g.,][]{marsden09,weiss09,austermann10,hatsukade11}, with
extremely wide-area surveys possible both from the ground
\citep{vieira10,marriage11}, and from space with the {\em Herschel
  Space Observatory}\footnote{{\em Herschel} is an ESA space
  observatory with science instruments provided by European-led
  Principal Investigator consortia and with important participation
  from NASA.}  \citep{pilbratt10}.  Large-area surveys are identifying
many very bright DSFGs whose fluxes are rivaled by those of only a few
extreme, serendipitously discovered objects that have been confirmed
to be gravitationally lensed \citep[e.g.,][]{swinbank10}.  Regardless
of the balance between intrinsically hyperluminous systems and less
extreme but gravitationally lensed galaxies within these samples, we
are no longer missing the rarest DSFGs because of limited sky
coverage. {\em Herschel} is playing a particularly important role
because the wavelength coverage of its SPIRE instrument
\citep{griffin10} allows selection of DSFG samples that are relatively
free of dust temperature biases \citep{magdis10}, although they are
limited by confusion to only the most extreme luminosity systems at $z
> 1$ \citep{symeonidis11}.  Specialized instruments designed for
wide-band spectral line surveys now enable the determination of blind
CO redshifts \citep[see, e.g.,][]{zmachines07, weiss09} for bright
DSFGs, without intermediate radio continuum mapping or optical
spectroscopy.  An example of the combination of these new developments
is the recent use of two wide-bandwidth instruments to obtain CO
redshifts for a complete sample of five bright {\em Herschel} sources
\citep{lupu10,frayer11}, an essential step in confirming their status
as galaxy-galaxy lenses \citep{negrello10}.

Expanding on this initial work, we here report on $\lambda \sim 1$\,cm
spectroscopy of the $^{12}$C$^{16}$O $J = 1-0$ ground-state rotational
transition toward two dozen of the brightest DSFGs in catalogs from
the {\em Herschel} Astrophysical Terahertz Large Area Survey
\citep[H-ATLAS; ][]{eales10, ibar10, pascale11, rigby11, smith11}
program, using the Zpectrometer ultra-wideband spectrometer on the
National Radio Astronomy Observatory's 100-meter diameter Robert C.\
Byrd Green Bank Telescope (GBT).  Submillimeter continuum flux ratios
provide some coarse redshift information for many sources, but the
precise redshifts needed to enable additional observations with
narrow-band instruments require spectroscopy of atomic or molecular
lines.  The Zpectrometer is one of several ultra-wideband
spectrometers built for this purpose, and is the first instrument to
make routine measurements of the CO $J=1-0$ rotational transition from
high-redshift galaxies \citep[e.g.,][]{swinbank10, harris10,
  negrello10, frayer11, scott11, riechers11a, riechers11b}. With the
CO molecule's $J = 1$ level only 5.4\,K above the ground state, its
low but nonzero permanent dipole moment, and its strong C--O bond, the
$J = 1-0$ transition is the best tracer of molecular gas over a wide
range of conditions in molecular clouds.  In addition to giving a
spectroscopic marker for redshift measurements, velocity-resolved
spectroscopy yields dynamical information, which together with gas
masses derived from CO intensities provides key inputs to
understanding DSFGs' star formation efficiencies and overall
evolutionary states.

Subsequent sections of this paper describe the observations and the
initial results from this sample.  Section~\ref{sec:obs} describes
target selection and observations, Section~\ref{sec:res} contains
observational results, and Section~\ref{sec:disc} provides further
analysis and discussion.  Calculations use a $\Lambda$CDM cosmology
with $\Omega_m = 0.27$, $\Omega_\Lambda = 0.73$, and $h_0 = 0.71$
\citep{spergel07}.

\pagebreak
\section{Observations}\label{sec:obs}
We drew our targets from early {\em Herschel}-ATLAS catalogs (H-ATLAS
collaboration, priv.\ comm.\ 2010) of continuum detections in the {\em
  Herschel} SPIRE instrument's 250, 350, and 500\um\ wavelength
photometric channels.  For this initial study, we selected bright
``350\um\ peaker'' galaxies with flux densities $S_\nu(350\,\mu{\rm
  m}) \geq 115$\,mJy and observed spectral energy distributions (SEDs)
peaking in the SPIRE 350\um\ band, within errors.  The catalogs
screened out local spiral galaxies and high-$z$ blazars with the
methods described in \citet{negrello07} and \citet{negrello10}.  This
provided a set of bright targets with peak far-infrared emission near
100\um\ wavelength in the rest frame for galaxies with $z \approx 2$
to 3.  Most of the targets were in the equatorial multi-wavelength
{\rm Galaxy And Mass Assembly} (GAMA) survey \citep{driver09} fields
near Right Ascension 9, 12, and 15 hours, with a few targets from the
North Galactic Pole (NGP) field near $\alpha = 13$ hours and $\delta =
27^\circ$.  Table~\ref{tab:obs} is a list of target positions and
integration time information.
  
\begin{deluxetable}{lrrrrr} 
\tabletypesize{\scriptsize}
\tablecaption{Target list grouped by observed pairs (see
  Sec.~\ref{sec:obs}). Positions are encoded in the source names
  following the IAU convention of providing Right Ascension and
  Declination, here in J2000.0 coordinates.  Other columns give the
  GAMA or North Galactic Pole (NGP) field, total integration time,
  number of individual observing sessions, and a target reference
  number for figures and other tables; letters denote targets with
  redshifts from the H-ATLAS Science Demonstration Phase
  \citetext{\citealp{negrello10}; $a$ is ID.17b, see also
    \citealp{lupu10}; $b$ is ID.130, see also \citealp{frayer11}}.
  Emission from the first source in each pair would appear in the
  positive sense in the spectra in Figure~\ref{fig:spectra}, and
  emission from the second source in the negative sense.  Positions
  and therefore names of sources in the NGP field are preliminary and
  may change slightly.  \label{tab:obs}}
 
\tablewidth{0pt} 
\tablecolumns{6}
\tablehead{ \colhead{H-ATLAS} & \colhead{Field} & \colhead{$t_{int}$}
  & \colhead{No.} &
  \colhead{Target}\\
  \colhead{} & \colhead{} & \colhead{$[$hr$]$} & \colhead{sess.} &
  \colhead{No.}  } \startdata
J083051.0+013224 & GAMA09 & 7.87 & 3 & 1\\
\vspace{6pt}
J084933.4+021443 &  &  &  & 2\\

J083929.5+023536 & GAMA09 & 7.08 & 3 & 3\\
\vspace{6pt}
J084259.9+024958 &  &  &  & 4\\ 

J090302.9-014127 & GAMA09 & 5.46 & 2 & $a$\\
\vspace{6pt}
J091305.0-005343 &  &  & & $b$\\

J091840.8+023047 & GAMA09 & 4.59 & 2 & 5\\
\vspace{6pt}
J085111.7+004933 &  &  &  & 6\\ 

J091948.8-005036 & GAMA09 & 3.67 & 2 & 7\\
\vspace{6pt}
J092135.6+000131 &  &  &  & 8\\ 

J113526.3-014605 & GAMA12 & 5.64 & 2 & 9\\   
\vspace{6pt}
J113243.1-005108 &  &  &  & 10\\     

J113833.3+004909 & GAMA12 & 3.67 & 2 & 11\\
\vspace{6pt}
J113803.5-011735 &  &  &  & 12\\ 

J114637.9-001132 & GAMA12 & 7.35 & 3 & 13 \\ 
\vspace{6pt}
J115112.3-012638 &  &  &  & 14\\  

J115820.2-013753 & GAMA12 & 5.25 & 2 & 15\\
\vspace{6pt}
J114752.7-005832 &  &  &  & 16\\ 

J132426.9+284452 & NGP & 2.89 & 2 & 17\\
\vspace{6pt}
J133008.3+245860 &  &  &  & 18\\

J134429.4+303036 & NGP & 3.94 & 2 & 19\\
\vspace{6pt}
J133649.9+291801 &  &  &  & 20\\

J141351.9-000026 & GAMA15 & 10.23 & 4 & 21\\
\vspace{6pt}
J142751.0+004233 &  &  &  & 22
\enddata
\end{deluxetable}

We observed the targets with the Zpectrometer analog lag
cross-correlator spectrometer \citep{zp07etal, harris05} connected to
the GBT's facility Ka-band receiver, which was configured as a
correlation receiver.  Receiver improvements in Fall 2010 extended its
high frequency performance, allowing spectroscopy from 25.6 to
37.7\,GHz, corresponding to the CO molecule's 115.27\,GHz $J = 1-0$
transition at redshifts of $2.1 \leq z \leq 3.5$.  The Zpectrometer's
spectral resolution is a sinc function with full-width at half maximum
({\rm FWHM}) of 20\,MHz, sufficient to provide a few resolution
elements across typical galaxy lineshapes.  Given the instrument's
wide overall bandwidth, the spectral resolution varied with frequency
across the spectra from 234 to 157\kms, and the {\rm FWHM} beamsize
from 27 to 16 arcsec.

A combination of the correlation receiver's ability to difference
power between two beams on the sky and the Zpectrometer's large
bandwidth allows different observing techniques from those common in
narrowband total-power radio astronomy.  Our instrument, observing
technique, data reduction, and calibration methods are fully described
in \citet{harris10}.  Briefly, the correlation receiver implementation
electrically differences power between the receiver's two input horns,
which are separated by 78\,arcsec on the sky.  We switched the source
between the two horns by moving the GBT's secondary mirror 78\,arcsec
in a 10\,s cycle, then differencing spectra from the two positions to
obtain a source spectrum.  To eliminate the residual tens of mJy of
spectral baseline structure from optical offsets, we observed pairs of
targets close in position on the sky, cycling between sources every
4\,min, again differencing this pair.  In this difference spectrum of
the two positions an emission line from the first source would appear
in the positive sense, while one from the second source would appear
in the negative sense.  Differencing left little optical offset and
baseline structure, at the cost of eliminating information about
individual source continua.

Observations were conducted in sessions of 3 to 8 hours duration (see
Table~\ref{tab:obs}) on dates from 2010 November through 2011 April
for a total of 64.5\,hr of on-sky observing time.  Counting observing
overheads, the actual elapsed observing time was about 130\,hr.  We
observed in a variety of weather conditions, mostly with reasonable to
good Ka-band atmospheric transmission and low wind.

We established absolute flux scales across the spectra by dividing the
astronomical source difference spectra by the spectrum of a bright
(few Jy) quasar suitable as a pointing reference near each target
pair.  In some cases we could use one of the cm-wave flux standards
3C48, 3C286, or 3C147 directly (0.80, 1.83, 1.47\,Jy at 32\,GHz,
respectively; \citeauthor{AstAlm2011} 2011), but for most of our
sources we cross-calibrated spectra of the nearby pointing source with
one or more of the standards.  We cross-checked 3C48's flux density
from \citeauthor{AstAlm2011} against a Mars flux density model by
B.~Butler\footnote{http://www.aoc.nrao.edu/$\sim$bbutler/work/mars/model/}
and found agreement within 2\%.  Comparison with the recommended 2012
January Jansky-VLA flux density for 3C286 is 1.96\,Jy (7\% higher)
than the value we use, and we find a Ka-band spectral index of $\alpha
= -0.8$ instead of the Jansky-VLA's $\alpha = -0.4$.
Cross-calibrations of the flux density for the quasar we used for the
GAMA09 sources against all three of the standards agreed within 10\%.
We determine calibrator spectral indices across the Ka-band, which can
be important in transferring the 32\,GHz standard fluxes to other
frequencies in the band, by comparison with Mars' blackbody $\alpha =
+2$.

We pointed and focused every hour on the bright quasars we had
selected as secondary flux calibrators near our sources.  Pointing
offsets were always within a third of a beam near band center.  We
took spectra of the pointing source at the beginning and end of each
pointing cycle to measure changes in optical gain and atmospheric
transmission.  Dividing the difference spectrum of an astronomical
source pair by the average spectrum of the pointing source not only
corrected for bandpass gain but also for the effects of pointing
errors and changing atmospheric transmission with timescales
comparable to an hour, or longer.  Session to session repeatability
for the quasar flux densities was generally within 10\%, and we take
20\% as representative of the overall amplitude calibration
uncertainty to include gain effects from pointing drift.

Version D of the Zpectrometer's standard GBTIDL scripts provided
quick-look assessment during observations and produced files for
combining data from different sessions.  We used version 5.4 of our
Zred package, which is written in the R language \citep{Rref}, for
further analysis.  The receiver produced systematic spectral structure
with fluctuating amplitude across many of the spectra.  Spectral lines
from galaxies were much narrower than features in the systematic
structures, which enabled us to remove the instrumental artifacts
\citetext{\citealp{harris10} contains further information}.  We removed
a systematic ripple with Fourier filtering and a complex but
fixed-pattern structure by fitting and subtracting median-filtered
session-average templates of the structure.  We removed no further
baseline structure, and with these corrections we could keep nearly
all of the data.

\subsection{Line detection algorithm \label{sec:lineDetAlg}}
In this exploratory phase of a full survey, we set integration times
to identify bright CO sources rather than to get high signal-to-noise
ratios on individual sources.  Since system noise changes across the
Zpectrometer's 12.1\,GHz band and nonideal noise is clearly present,
attempting to estimate noise for signal-to-noise ratios by computing
the channel fluctuations across the spectrum's entire frequency range
would overestimate noise in some parts of the spectrum and
underestimate it in others.  Instead, we harness time-series
information from sub-integrations to make individual channel noise
estimates as part of a detection confidence algorithm.
\citet{harris10} has a complete discussion of this algorithm, which
analyzes whether the amplitude in each frequency channel is
statistically higher than the average of its neighbors, based on a
generalized Student-$t$ test of many bootstrapped realizations of each
spectrum.

The algorithm's main advantage over traditional methods is that it
treats noise in individual frequency channels.  Its main weakness
stems from its assumption that the underlying spectral baselines are
relatively smooth on small scales so that local averages are
representative.  Failure of this assumption keeps the computed
confidence level from being absolute, as discontinuities in spectral
baselines and small fluctuations near bright features can all register
as potential detections.

Our experience with the algorithm on Zpectrometer data is very good,
based on comparisons of tentative detections within subsets of data
and from detections of different lines from the same galaxies with
other instruments.  Given deviations from the assumptions caused by
nonideal noise of a few hundred $\mu$Jy, the algorithm remains a
powerful guide rather than an exact indicator for identifying weak
lines.  Experience gained from viewing many spectra toward different
sources under different weather conditions, but all with the same
frequency coverage, provides valuable impressions of typical baseline
structures and noise behavior, further aiding in screening against
spurious detections.  All detections here come from at least two
independent observing sessions, and most have at least tentative
detections in individual sessions.  By generating 1000 bootstrapped
spectra as part of the detection algorithm, the data are thoroughly
mixed by time and session.  Sweeping through a wide range of binnings
ensures that a detection is not based on a single favorable choice.
We have been conservative in our decisions, while recognizing that a
small number of false positives are more beneficial than false
negatives in observations designed to spur followup work.  In
practice, independent observations of other lines from the same
targets have justified our detection selections.

\section{Results}\label{sec:res} Panels in
Figure~\ref{fig:spectra} show the spectra and confidence plots for all
of the source pairs we observed.  The upper panel for each source pair
is the spectrum across the Zpectrometer band.  Vertical dashed lines
mark line frequencies of detected galaxies.  The lower panel shows the
detection probabilities $p$ from our detection algorithm, versus
frequency, given as Confidence\,$=-\log_{10}(1-p)$ (numerically, the
scale is equivalent to the number of nines in confidence: 0.9, 0.99,
0.999, etc.\ for $p \geq 0.9$).  Each dot in the plot is an individual
channel's confidence measure (within the algorithm's assumptions) for
a given combination of binning width and starting point.  Columns of
dots at specific frequencies show where a potential detection is
relatively immune to exact binning parameters, indicating that a real
line is present rather than a favorable binning for a chance
fluctuation.

\renewcommand{\thefigure}{1}
\begin{figure*}
\centering \includegraphics[height=6.5in, clip=true, angle=-90]{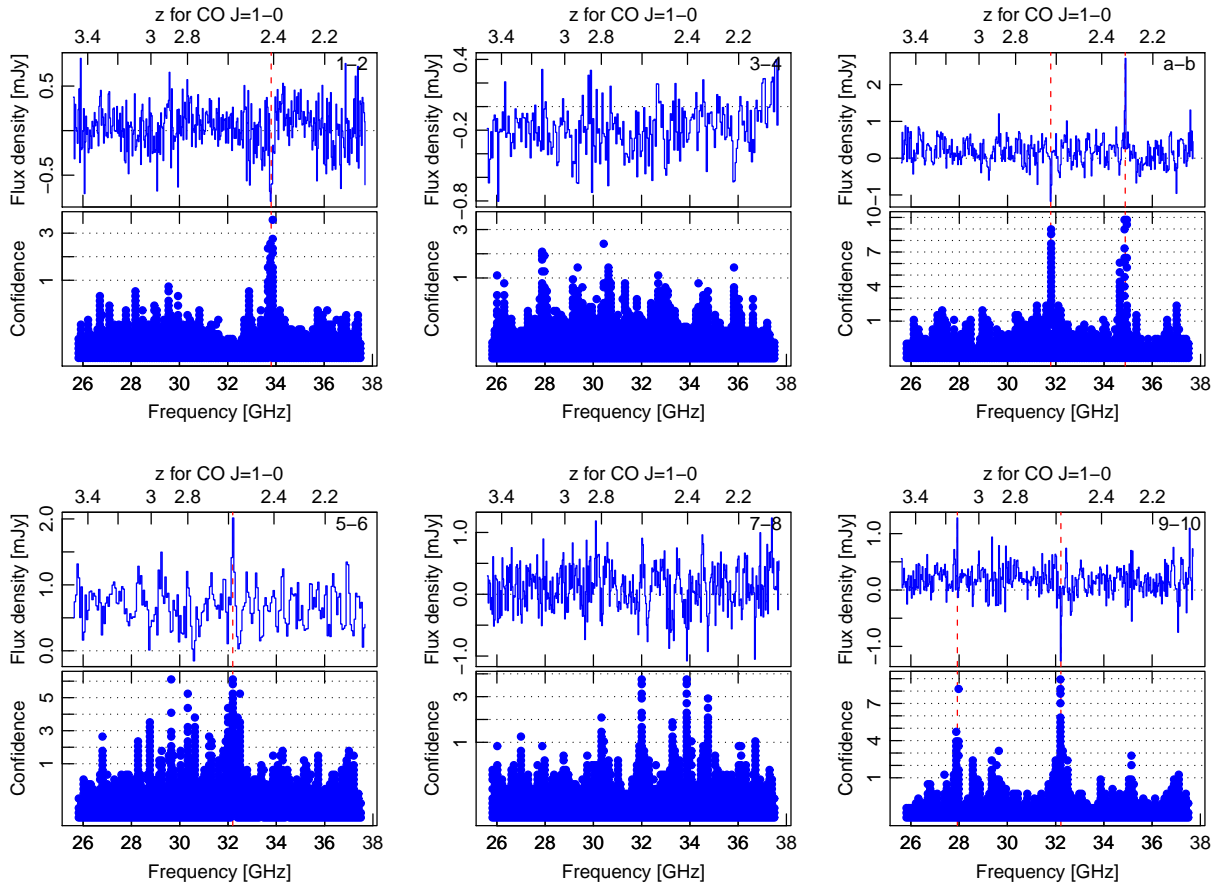}
\caption{The first twelve sets of spectra and detection confidence plots for
  targets labeled by target number in Table~\ref{tab:obs} and described in
  Section~\ref{sec:res}.  The upper
  panel in each set is the
  difference spectrum for each pair (panel letter in the upper right
  hand corner).  The lower panel is a detection confidence plot as 
  described in the
  text.  Coherent columns of dots indicate detections that are
  insensitive to binning parameters.  Vertical dashed lines mark
  detections or tentative detections.  
\label{fig:spectra}}
\end{figure*}
\renewcommand{\thefigure}{\mbox{1 (cont.)}}
\begin{figure*}
\centering \includegraphics[height=6.5in, clip=true, angle=-90]{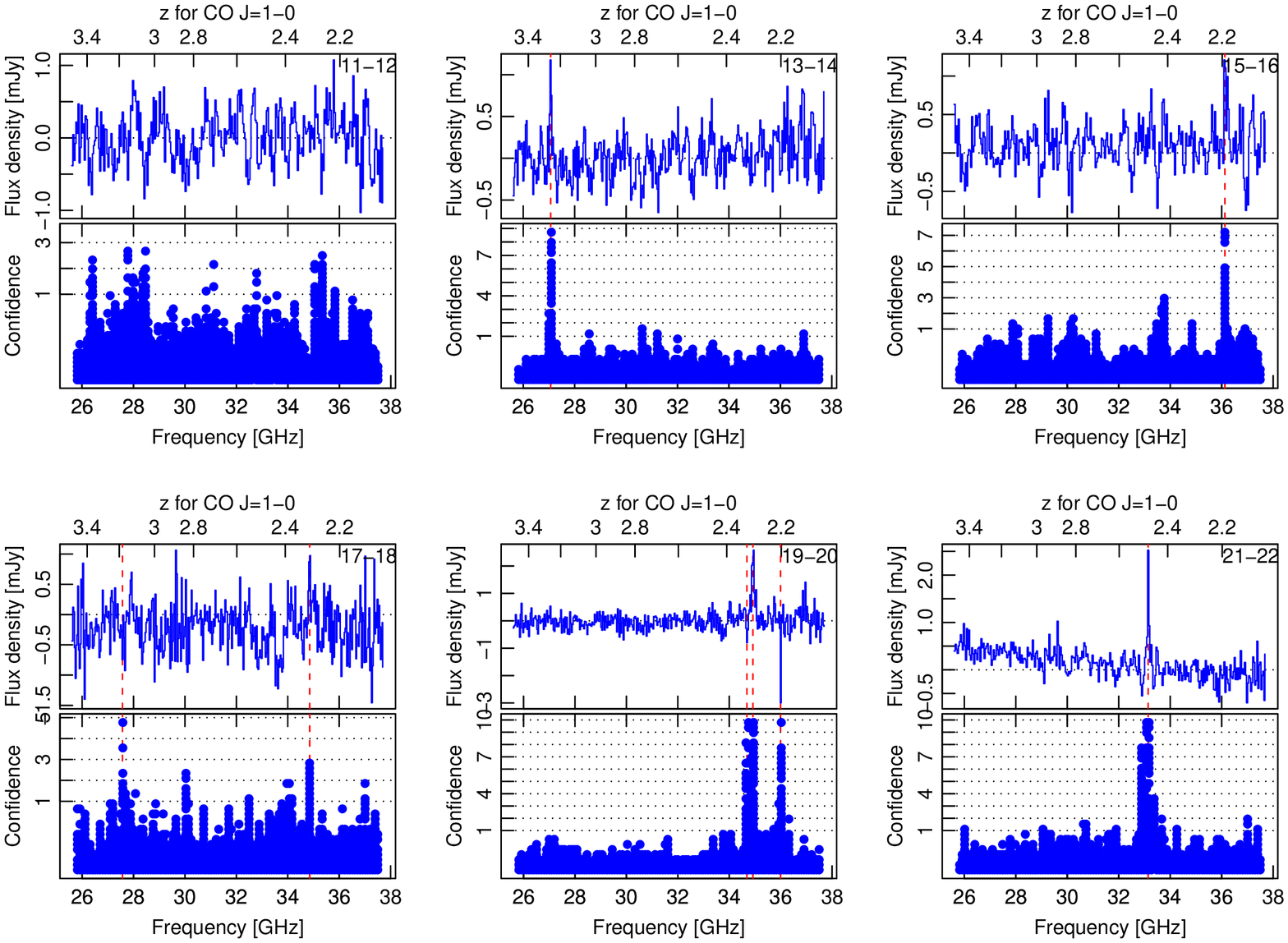}
\caption{The second twelve sets of spectra and detection confidence plots for
  targets labeled by target number in Table~\ref{tab:obs} and described in
  Section~\ref{sec:res}.
}
\end{figure*}
\renewcommand{\thefigure}{\arabic{figure}}
\setcounter{figure}{1}

Our brief comments on individual spectra are: 

\noindent{\em J083051.0+013224} and {\em J084933.4+021443}
(Fig.~\ref{fig:spectra}, 1-2): Only the second target in the pair
(emission appears in the negative sense) is detected.
D.A.~Riechers (priv.\ comm.\ 2011) has detected a single
strong line with the CARMA observatory toward the first in the pair.
If the line were CO $J = 3-2$, J083051.0+013224 would lie at a
redshift with the $1-0$ line in a low-noise region of the Zpectrometer
band.  This is a statistically significant nondetection that we discuss in
section~\ref{sec:completeness}.

\noindent{\em J083929.5+023536} and {\em J084259.9+024958}
(Fig.~\ref{fig:spectra}, 3-4): Neither target is clearly detected in this
spectrum.

\noindent{\em J090302.9$-$014127} and {\em J091305.0$-$005343}
(Fig.~\ref{fig:spectra}, a-c): Clear detections of both targets.  The
high confidence at slightly lower frequency than the strong positive
line is likely an artifact, as even a modest dip can be far from the
local amplitude mean, a signature the detection algorithm interprets
as a line.

\noindent{\em J091840.8+023047} and {\em J085111.7+004933}
(Fig.~\ref{fig:spectra}, 5-6): Clear detection of the first target in the
pair.  The continuum offset from zero shows that the first
target has a higher continuum flux than the second target.  

\noindent{\em J091948.8$-$005036} and {\em J092135.6+000131}
(Fig.~\ref{fig:spectra}, 7-8): Neither target is clearly detected in this
spectrum.  Residual large-scale structure may obscure what could be
tentative detections.  Even small noise fluctuations at the tops of large-scale
positive and negative structures are far from local means, so they
register strongly in the confidence plot.

\noindent{\em J113526.3$-$014605} and {\em J113243.1$-$005108}
(Fig.~\ref{fig:spectra}, 9-10): Clear detections of both targets.
Some spurious high confidence peaks are associated with each of the
bright lines.

\noindent{\em J113833.3+004909} and {\em J113803.5$-$011735}
(Fig.~\ref{fig:spectra}, 11-12): Neither target is clearly detected in this
spectrum.

\noindent{\em J114637.9$-$001132} and {\em J115112.3$-$012638}
(Fig.~\ref{fig:spectra}, 13-14): Clear detection of the first target.

\noindent{\em J115820.2$-$013753} and {\em J114752.7$-$005832}
(Fig.~\ref{fig:spectra}, 15-16): Clear detection of the first target.

\noindent{\em J132426.9+284452} and {\em J133008.3+245860}
(Fig.~\ref{fig:spectra}, 17-18): Tentative detection of the first
target.  Baseline structure to slightly higher frequencies makes it
difficult to find a local baseline, so the line parameters are
uncertain.

\noindent{\em J134429.4+303036} and {\em J133649.9+291801}
(Fig.~\ref{fig:spectra}, 19-20): Clear detections of both targets,
with one line in the positive sense and two in the negative sense.  We
attribute the stronger negative line with the SPIRE source.  For
completeness, we assign the other negative line a tentative formal
detection because it is uncharacteristically broad for a spurious
signal, but it is close to the positive-sense line where the local mean
changes rapidly and may throw off the detection algorithm.

\noindent{\em J141351.9$-$000026} and {\em J142751.0+004233}
(Fig.~\ref{fig:spectra}, 21-22): A strong detection of the first target.
The high confidence measures for dips to either side are most likely due
to a high local mean in the region.  The dips are rather wide to be due to a
galaxy, so it is unlikely that they represent detections.  The
continuum slope indicates a difference in spectral index between the
two targets.

\begin{figure*}
\centering 
\includegraphics[height=7in, clip=true, angle=-90]{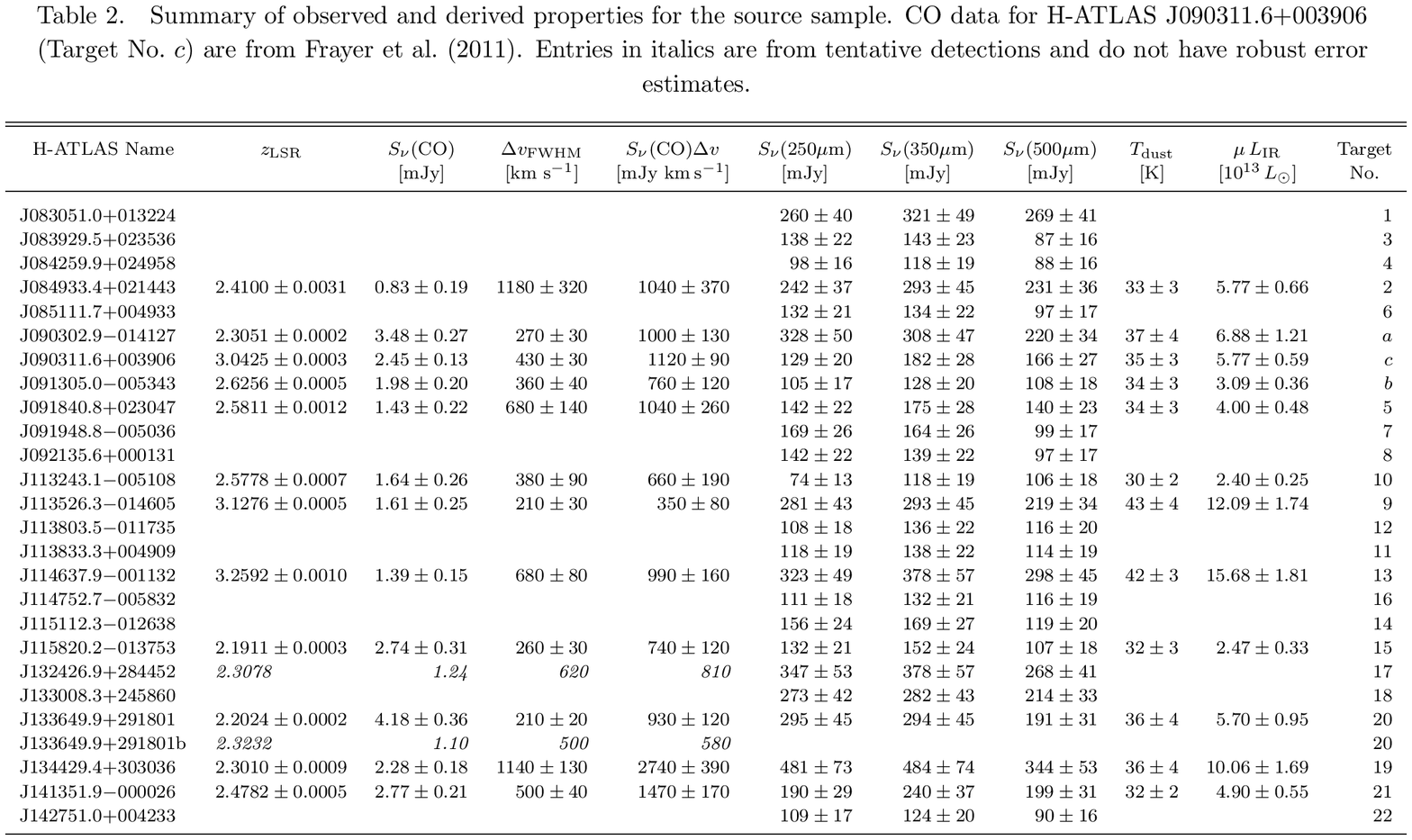}
\label{tab:summary}
\end{figure*}

Overall, we detected 11 of the 24 targets in our sample. Two of the
detections were blind independent confirmations of sources with
established redshifts \citetext{ID.17b in \citealp{lupu10} and ID.130
  in \citealp{frayer11}}.  This success rate is similar to that of
detections from the Plateau de Bure millimeter-wave interferometer
starting from optical redshift catalogs \citep[e.g.,][]{neri03,
  greve05, tacconi06, bothwell11}, but without the complications
associated with finding optical redshifts for submillimeter sources
\citep[see, e.g.,][]{chapman05}.  Section~\ref{sec:completeness}
contains a more extensive discussion of detection completeness.  In
addition to the 11 detections, we also list two tentative detections,
denoted by italics in Table~\ref{tab:summary} and (for
J132426.9+284452) with open circles in the figures.  Line parameters
for tentative detections were too uncertain for robust error estimates.

Based on the submillimeter photometric selection and line strength, we
initially derived redshifts assuming that the lines were the
redshifted CO $J = 1-0$ transition rather than CO $J = 2-1$ from a
$5.1 \leq z \leq 8.0$ galaxy or a line from a species other than CO.
This assumption has proved correct for all 11 sources, which have
other observed lines, most starting with redshifts from the
Zpectrometer observations \citep[D.A.\ Riechers, priv.\ comm.\ 2011;
P.P.\ van der Werf, priv.\ comm.\ 2011;][]{lupu10}.  

Table~\ref{tab:summary} summarizes observed and derived source
parameters.  For detected lines, the parameters are from
single-component fits to Gaussian lineshapes, with errors given by the
statistical uncertainties in the fit at the 68\% (``$1 \sigma$'')
confidence level.  Gaussian fits to the convolutions of Gaussian line
shapes and the correlator's sinc instrumental profile shows that
linewidth corrections are unimportant for linewidths above 200\kms\
FWHM (the correction is 12\% when the Gaussian line and sinc FWHMs are
equal, falling below 1\% when the Gaussian's FWHM is 1.5 times the
sinc's FWHM or wider).  All lines are broader than this, so we make no
corrections.  The table also contains estimates of the total infrared
(8--1000\um) fluxes and dust temperatures obtained from fits to the
SPIRE photometry (H-ATLAS collaboration, priv.\ comm.\ 2010), as
discussed later.  In addition to the detections from this program,
Table~\ref{tab:summary} also includes CO $1-0$ data from a previous
Zpectrometer detection of H-ATLAS J090311.6+003906
\citep[][ID.81]{frayer11}.

\setcounter{figure}{1}
\setcounter{table}{2}

\subsection{Flux relationships and detection completeness  \label{sec:completeness}}

\begin{figure} 
\centering \includegraphics[width=3.2in, clip=true]{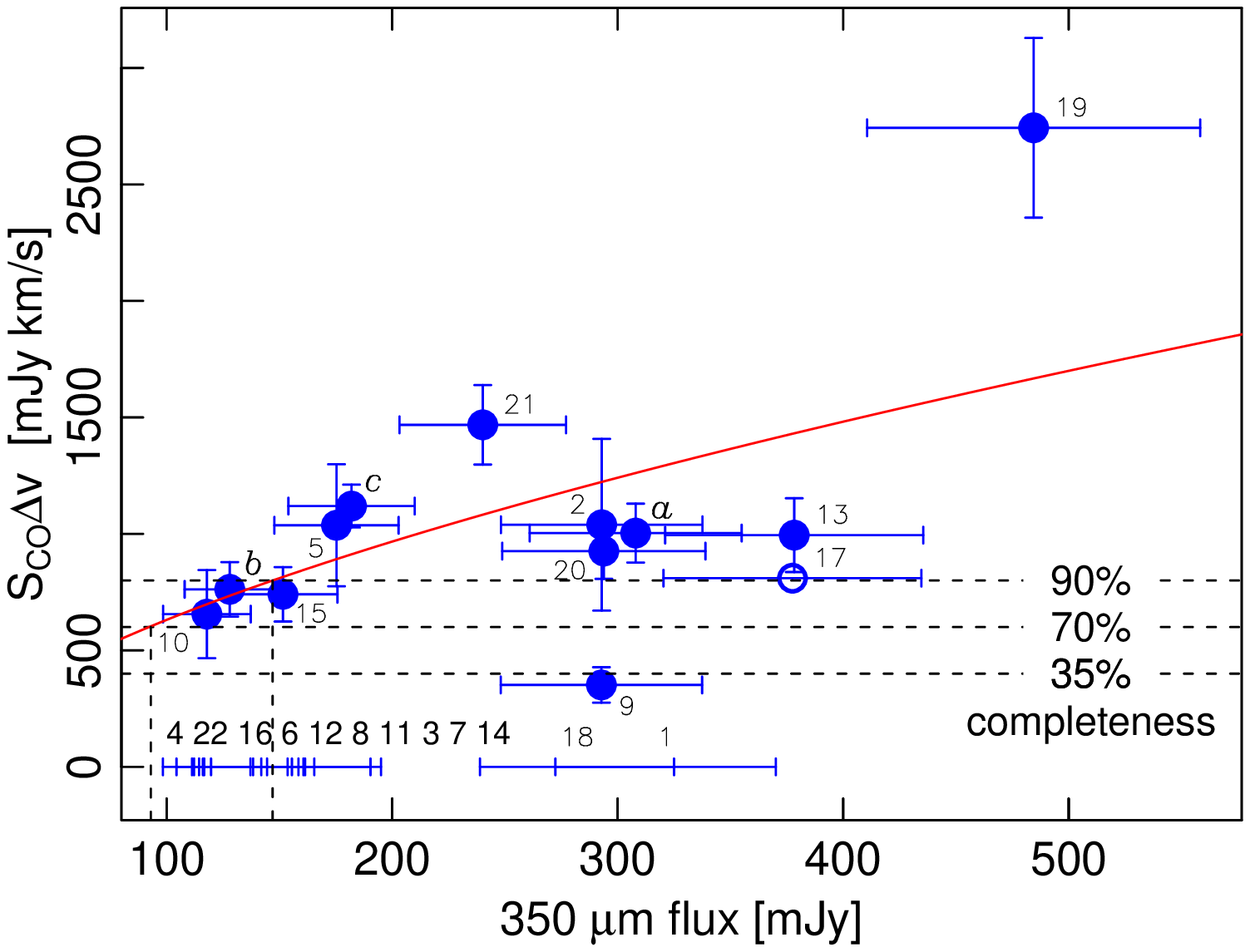}
\caption{Integrated CO flux \Sco\ versus 350\um\ flux density \Scont\  
for CO-detected and CO-undetected sources from our sample.  
  CO nondetections have been assigned zero flux.  
  Horizontal dashed lines show completeness levels for line detections derived 
  from our 400\kms\ linewidth simulations, with corresponding 
  \Scont\ derived from the power-law fit shown by the smooth
  curve. All galaxies with CO
  nondetections have 350\um\ flux densities above the 70\%
  completeness limit for detections if a simple CO-continuum flux
  scaling law holds
  (see text).  This implies that chance alone cannot explain the
  fraction of galaxies in the sample that we do
  not detect, but that some galaxies have redshifts outside the
  Zpectrometer band or contain relatively little molecular gas.
  Points are labeled by target 
  numbers given in the tables.
  \label{fig:ScoVsS350}}
\end{figure}

We have firm detections of 11 of the 24 targets in our program, a
number large enough to draw some sample conclusions.  Here we use the
molecular and continuum flux information to explore the detection
completeness and to evaluate possible reasons for CO nondetections.
Figure~\ref{fig:ScoVsS350} explores the relationship between 350\um\
flux density, \Scont, and the CO $J = 1-0$ integrated flux, \Sco.
Sources without CO detections have zero amplitude in this plot, and
the open circle denotes a tentative CO detection.  Line nondetections
generally fall to lower 350\um\ flux densities, but several galaxies
are bright in continuum but not detected in CO.  We can take the
third-brightest \Scont\ source in our sample (J083051.0+013224, target
number 1 in figures and tables) as an example of a nondetection that
implies a galaxy with a redshift outside the Zpectrometer band or an
abnormally low CO to continuum flux ratio.  The former seems more
likely in this case: as noted above, CARMA has a clear detection of a
mid-$J$ CO line from this source, but deep integrations by the
Zpectrometer and other instruments have not found other lines
corresponding to a redshift within the Zpectrometer's $z = 2$--3.5
band.

Most likely, some fraction of the nondetected galaxies have redshifts
that are not well predicted by continuum properties, while others have
CO $1-0$ lines that are fainter than our detection threshold.  Before
we simply ascribed CO nondetections to galaxies with observed
luminosities below some threshold, we first examined the line detection
completeness and the relationship between line and continuum fluxes.
Estimating line detection completeness in broadband spectra is
complicated by system temperature and nonideal noise that vary with
frequency; this frustrates any attempt to define a quantity such as
the baseline {\rm rms} in narrowband spectra that could specify a
simple detection limit.  We estimated completeness levels for
nondetections by simulation, adding sets of synthetic lines as
frequency combs across subscans for all sources, running the modified
spectra through the data reduction pipeline, and inspecting the
spectra to see which synthetic lines we could clearly identify.  With
a comb of seven 400\kms\ wide lines (a width close to the median of
the astronomical source linewidths) across each spectrum, we recovered
90\% of the synthetic lines with CO integrated intensity \Sco~=
800~mJy\kms, 70\% of the lines with \Sco~= 600~mJy\kms, and 35\% of
those with \Sco~= 400~mJy\kms.  Taking \Sco~= 600~mJy\kms as the
typical lower limit for our detections, the simulation gives 70\%
completeness for 400\kms\ lines, 90\% completeness for 200\kms\ lines,
and 40\% completeness for 800\kms\ lines.  This shows that the
statistical algorithm is most sensitive to peak intensity for lines
near detection thresholds.  The average completeness for these three
widths is 67\%, so 70\% is representative for a set of lines with
various widths and 600~mJy\kms.

A somewhat monotonic relationship between the continuum and line
fluxes in DSFGs must exist: energy balance requires that galaxies with
little far-IR continuum luminosity will have weak molecular emission.
The exact relationship is unknown, but a power-law fit established a
representative correspondence between \Sco\ and 350\um\ flux densities
\Scont\ for sources with CO detections.  This fit yielded equivalent
limits of \Scont~= 150\,mJy, 90\,mJy, and 50\,mJy for 90, 70, and 35\%
completeness for the median (400\kms) linewidth, as shown in
Figure~\ref{fig:ScoVsS350}.  Given the \Sco--\Scont\ distribution, the
exact form of the continuum--line relationship is unimportant over the
relatively small range at low flux densities, and a linear fit gave
essentially the same values.  Excluding the extreme high-flux point
from the fit flattens the \Sco--\Scont\ relationship, pushing the
completeness limits to lower continuum flux densities.  Pushing the
completeness limits to higher flux densities would require a steeper
relationship than could be supported by these data with a simple
model.

Whatever the exact form of the correspondence between \Sco\ and
\Scont, all of our targets have \Scont\ falling above the 70\%
completeness level (probability of detection) if their CO $J=1-0$
emission falls within the Zpectrometer frequency range.  We estimated
the number of targets we might expect to have missed due to faint CO
flux alone by considering the targets below \Scont~= 150\,mJy, the
approximate 90\% completeness level.  There are 10 sources in our
target list below this limit, of which we detect only two.  If chance
alone dominates, the detection rate will be given by the binomial
distribution.  Taking a lower limit of a 70\% detection probability,
the distribution finds 7 detections as most likely, with $7 \pm 3$
detections accounting for 99\% of the total probability.  The
probability of detecting just two sources is 0.1\%.  Increasing either
the detection probability or the flux limit corresponding to a given
detection probability reduces the probability of detecting just two of
ten sources.

If chance alone ruled, we should therefore have detected some 2 to 5
more weak sources. Considering the additional four nondetections with
350\um\ flux densities significantly above the nominal 90\%
completeness limit, this analysis points to a strong disparity between
actual and expected detections.  We conclude that chance is not the
only reason for nondetections, but that systematic effects are also
important: some of the sources could not be detected because their
redshifts are outside of the Zpectrometer's band, some because they
have lower \Sco/\Scont\ ratios than the detected sources, some because
their lines are weak and broad (although there is no sign of these
even though the eye is good at picking out correlated channels), or
some combination.

\subsection{Redshift distribution}
\begin{figure}  
\centering \includegraphics[width=3.2in, clip=true]{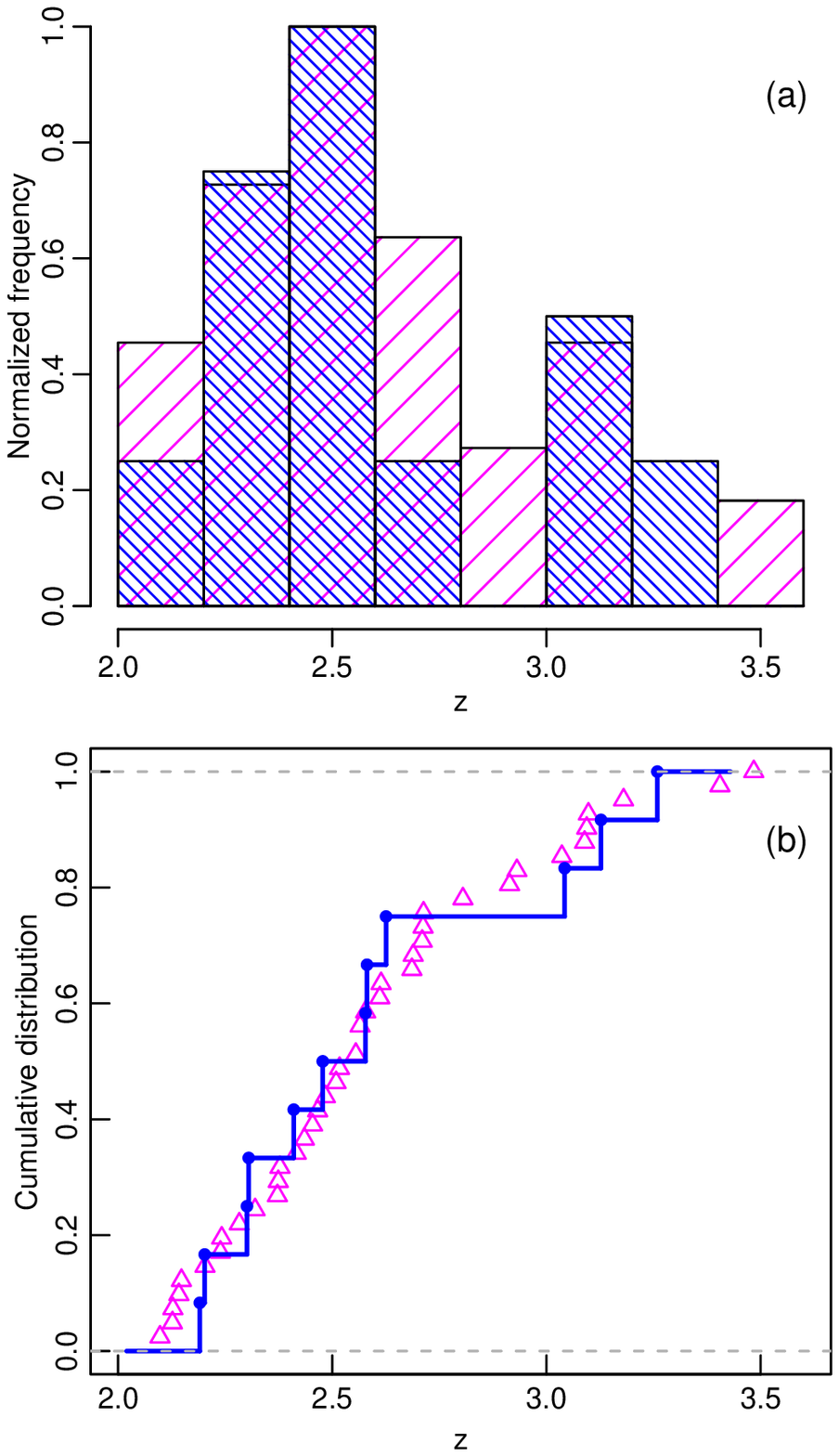}
\caption{Comparisons of the redshift distribution of galaxies from our
Zpectrometer CO survey and those from the 
  \citet{chapman05} optically-derived redshift survey of SMGs that 
  fall within the
  Zpectrometer's redshift range.  Panel (a) is the binned
  distribution with Zpectrometer data in close hatch and
  \citet{chapman05} data in coarser hatch.  Panel (b) gives the
  cumulative distribution functions, which do not rely on bin width
  choices; Zpectrometer data are solid points connected by lines,
  \citet{chapman05} data open triangles.  The two distributions are
  strikingly similar (a Kolmogorov-Smirnov test  returns a probability of 0.995
that they are drawn from the same parent population)
  although the selection criteria are completely different.  
This suggests a common parent population (and perhaps selection effects)
for the SPIRE- and 870\um-selected galaxies.
\label{fig:zDistribs}}
\end{figure}

Zpectrometer redshifts for the H-ATLAS sources provide an independent
sample for comparison with previous spectroscopic redshift surveys.
Comparing surveys, we find that the redshift distributions of the
``350\um\ peaker'' galaxies with Zpectrometer detections and those of
the radio-preselected SMGs \citep{chapman05} that fall within the
Zpectrometer's redshift range are strikingly similar, although the
source selection and lines used in the redshift measurements were
quite different: our galaxies are 350\um-bright targets from a
wide-area survey that highlights lensed sources, while the
\citet{chapman05} galaxies must be bright at 850\um\ and 20\,cm radio
continuum.  Figure~\ref{fig:zDistribs}a gives the binned
distribution for the galaxies with Zpectrometer CO detections.  The
median of the Zpectrometer CO redshifts is $z = 2.47 \pm 0.11$, and
the mean is $z = 2.60 \pm 0.10$ (68\% confidence levels by bootstrap
analysis), both well below the band center at $z = 2.78$.
Figure~\ref{fig:zDistribs}a shows that the peak of the observed
density function is near $z \sim 2.3$, agreeing well with the peaks
found for optical spectroscopic redshifts of SCUBA-selected sources
\citep{chapman05}, and for photometric redshifts of both {\em
  Herschel}-selected sources bright at 350\um\ \citep{amblard10} and
870\um-selected sources identified by LABOCA \citep{wardlow11}.

Figure~\ref{fig:zDistribs}b gives the cumulative distribution
functions of the Zpectrometer and the \citet{chapman05} sample
redshifts, allowing a clean comparison free of binning effects.  It
shows that the distributions are indistinguishable; a
Kolmogorov-Smirnov test gives a probability of 0.995 that the two
underlying distributions are the same.  Such good agreement must in
part reflect random chance in a statistically small sample, but
nevertheless it is clear that the two distributions are very similar.
There is otherwise no {\em a priori} reason that the distributions
should be so similar: for instance, the \citet{chapman05} sample could
be concentrated to lower redshifts because of the radio pre-selection,
while the Zpectrometer detections could highlight a population of
galaxies with strong molecular but very little rest-frame UV line
emission.

Another similarity between the Zpectrometer sample and galaxies
detected in millimeter-wave CO followup from optical redshift catalogs
is the distribution of CO linewidths.  Linewidths trace the dynamics,
and thus to some extent masses, of the galaxies, and are not strongly
affected by lensing.  Full width at half maximum ({\rm FWHM})
linewidths in the Zpectrometer sample range from 210 to 1180\kms, with
a median width of $400 \pm 100$\kms\ (68\% confidence levels by
bootstrap analysis).  In general this is somewhat narrower than the
widths of mid-$J$ lines in the millimeter-wave studies of
\citet{greve05} and \citet{tacconi06}, but a permutation test shows
that the difference in median widths between the studies is
significant at only the 70\% level, and could easily be from small
number statistics.  Combining the samples, a characteristic width of
about 500\kms\ is representative for DSFGs.  In absolute terms, the
widest $1-0$ lines are very broad, however, including two with {\rm
  FWHM} linewidths greater than 1100\kms.  If tracing virialized
matter, these widths correspond to emission from very massive galaxies
or interacting massive galaxies.

\subsection{Continuum properties \label{sec:continuum}}
Spectroscopic redshift measurements unambiguously break the $T_D$-$z$
degeneracy \citep[e.g.,][]{blain99degen} that renders dust temperature
estimates uncertain for galaxies without firm redshift measurements.
We fit a simple optically thin single-temperature dust model with dust
emissivity $\beta = 1.5$ to the 250, 350, and 500\um\ SPIRE flux
densities.  To estimate the observed total infrared (rest-frame
8--1000\um) luminosity, we joined the far-infrared fit smoothly in
slope to a power-law spectrum with form $S_\nu \propto \nu^{-1.4}$ at
short wavelengths \citep{blain03}.  Assuming that the same
gravitational magnification $\mu$ applies to emission in all far-IR
wavebands, lensing should not affect temperature estimates, but it
will scale the intrinsic luminosity $L_{\rm IR}$ so the observed
infrared luminosity is $\mu L_{\rm IR}$.

\begin{figure} 
%\epsscale{0.6}
%\plotone{figs/dustVsAlla.eps} %fig4, dustVsAlla.eps
\centering \includegraphics[width=3.2in, clip=true]{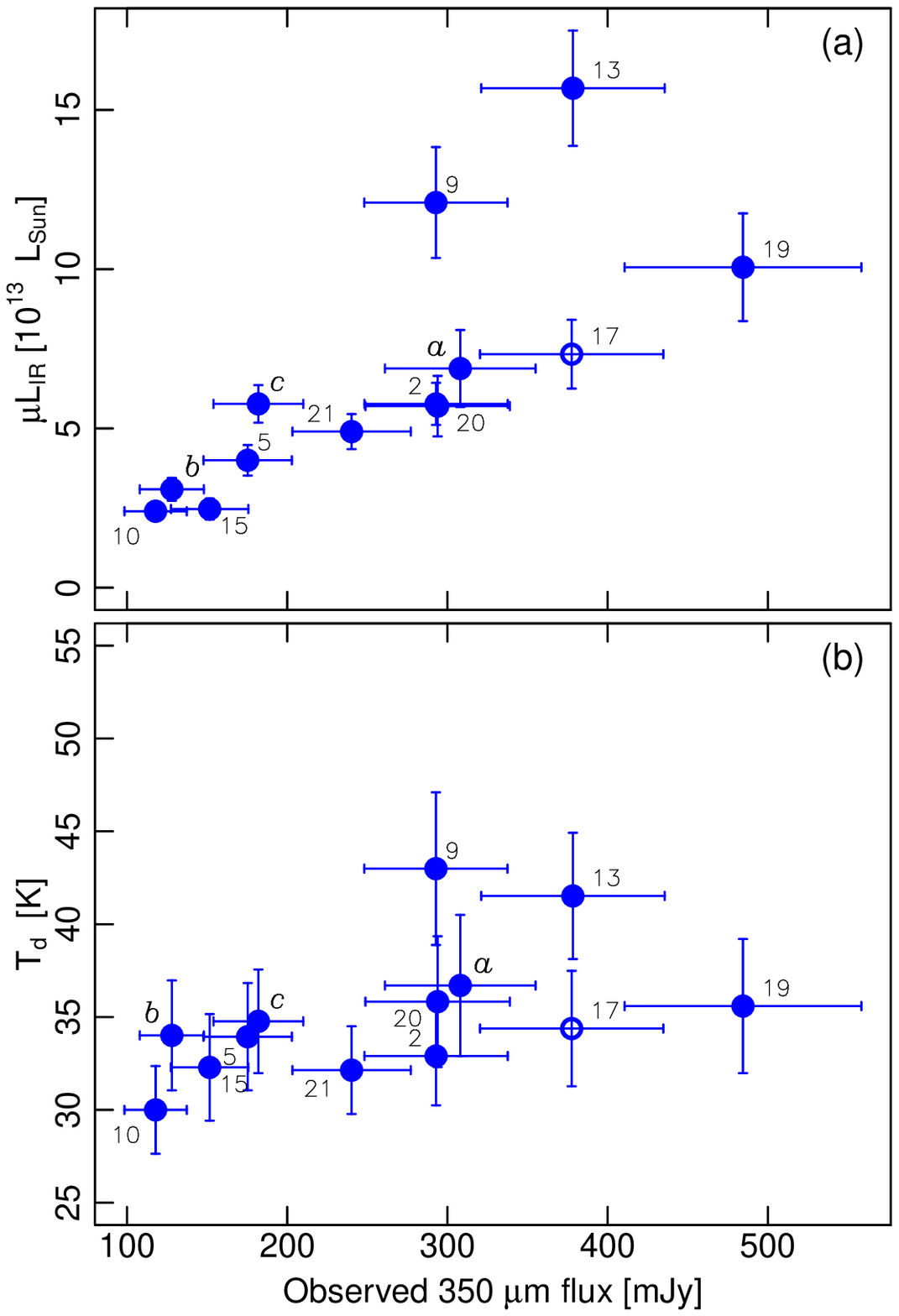}
\caption{Panel (a) gives the observed IR luminosity versus observed 350\um\ flux
  density \Scont, showing that the luminosity scales reasonably well with flux
  density, even in the presence of a range of
redshifts and dust temperatures.  Panel (b) shows that the derived
characteristic 
dust temperatures are nearly constant with \Scont,
  possibly at least in part from selection effects. The highest two points in
  both panels are from the two galaxies in our sample with $z \gtrsim 3.1$.
  Points are labeled by target numbers given in the tables.
  \label{fig:dustVsAlla}}
\end{figure}

Table~\ref{tab:summary} contains the model results and Monte-Carlo
error estimates for individual galaxies.  Excluding the two $z > 3.1$
galaxies, the dust temperature is $34 \pm 2$\,K averaged over all of
the detected sources, with a slight dependence on the observed 350\um\
flux density (Fig.~\ref{fig:dustVsAlla}b) or, equivalently, $\mu
L_{\rm IR}$.  Little temperature scatter is to be expected among the
detected galaxies because the source selection criteria favored
similar SPIRE-band SEDs.  Temperatures near 35\,K are similar to those
derived from other surveys of SMGs \citetext{$35\pm3$\,K,
  \citealp{kovacs06}; $34\pm5$\,K, \citealp{chapman10}; $37\pm1$\,K, 
  \citealp{wardlow11}}.  The two highest-$z$ galaxies in our sample have
temperatures of about 40\,K, a temperature higher than scatter alone
can explain.  Varying the dust emissivity parameter $\beta$ from 1.2
to 1.7 did not change the dust temperatures by more than 3\,K.  Dust
emission models that allow for dust emission optical depths yield
temperatures about 15\,K higher than those from the optically thin
model; we quote the optically thin results to facilitate comparisons
with previous work, most of which uses the same formalism.

Figure~\ref{fig:dustVsAlla}a shows that the 350\um\ flux density is an
excellent proxy for the observed infrared luminosity $\mu L_{\rm IR}$
at $z\sim 2$--3.  The two points falling above the general trend are
the two galaxies with the highest redshifts, $z > 3.1$, in our sample.
Excluding these two galaxies, we derive $\mu L_{\rm IR} = (S_\nu
(350\,\mu {\rm m})/(51 \, \rm{mJy})+ 0.4) \times 10^{13} \, L_\odot$.
Apart from the two galaxies at $z > 3.1$ there is no dependence of
$\mu L_{\rm IR}$ on redshift within errors.

\subsection{Comparison of photometric and line redshifts}
One of the goals of this project was to provide a data set suitable
for evaluating the precision of photometric redshifts against
spectroscopic measurements.  Here we make a first-cut comparison of
techniques based on simple SPIRE 3-band colors and on fitting to
template galaxy SEDs.

\begin{figure} 
\centering \includegraphics[width=3.2in, clip=true]{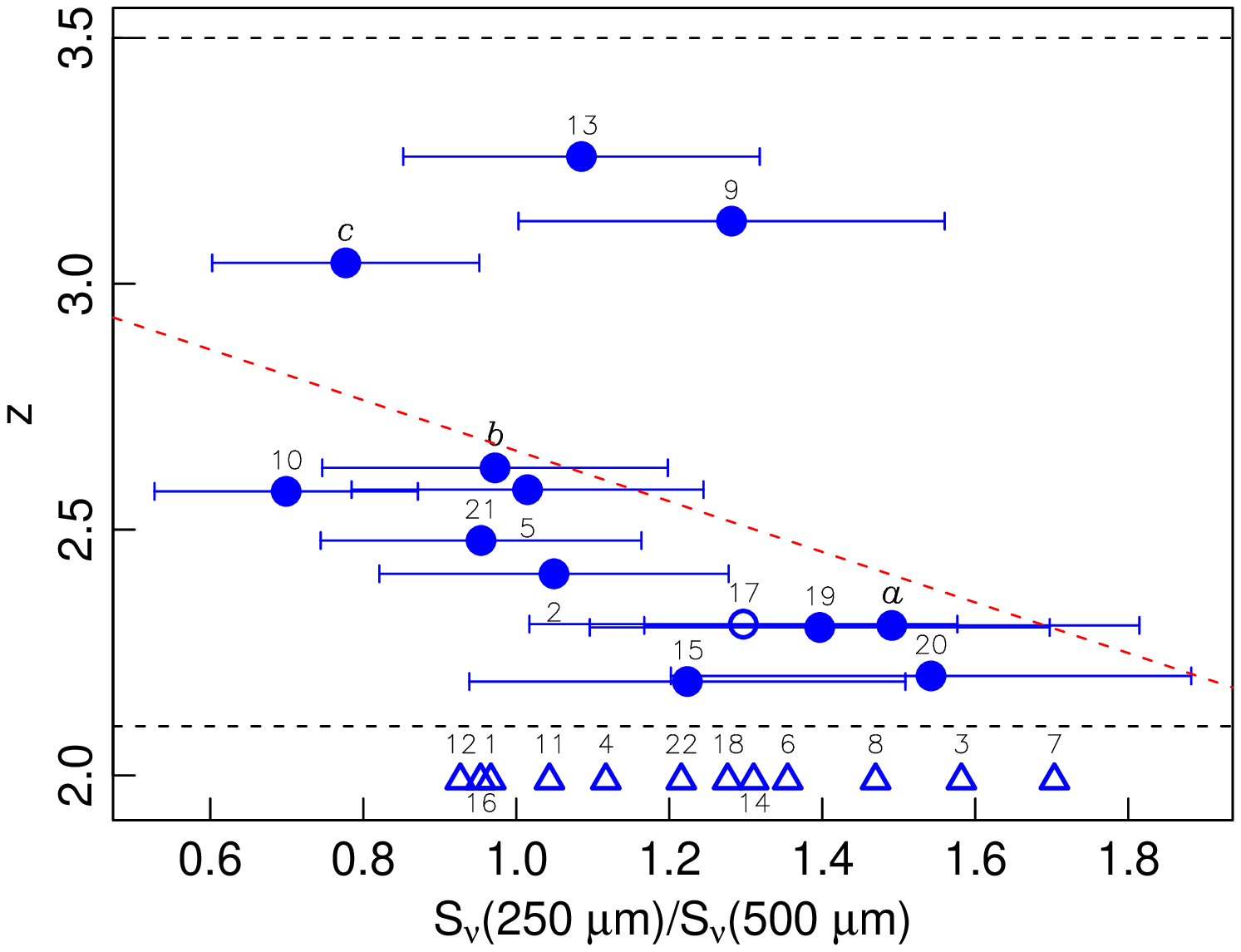}
\caption{Galaxy redshift $z$ versus 250\um/500\um\ continuum flux
  density ratio (solid points; open point for a tentative detection).  Dashed
  horizontal lines mark the Zpectrometer's band edges.  CO
  nondetections (triangles) are plotted outside the band at an
  arbitrary 
  $z = 2$.  For galaxies with CO
  detections, the figure shows that SPIRE colors give $\Delta z \sim
  0.3$ for this
  carefully filtered homogeneous sample.  Points are labeled by target 
  numbers given in the tables.
 \label{fig:ratioDetZ}}
\end{figure}

With galaxies chosen as ``350\um\ peakers,'' the simplest color
selection is a flux ratio that estimates the SED's peak wavelength.
Since the 250\um\ and 500\um\ bands straddle the emission peak their
ratio provides this estimate, reflecting some combination of
temperature and redshift.  As section~\ref{sec:continuum} shows, our
sample of galaxies has a small range of dust temperatures, so we can
test whether there is a simple relationship between the observed SED
peak wavelength and redshift.  Figure~\ref{fig:ratioDetZ} shows that
there is a relationship between the peak wavelength, as given by
the 250\um/500\um\ flux density ratio, and spectroscopic redshift for
galaxies with CO detections.  For these galaxies the 250\um/500\um\
flux density ratio predicts redshifts within $\Delta z = 0.3$
(standard deviation of the redshift error).  Other combinations of
continuum flux densities and ratios are less successful at predicting
redshifts than the peak wavelength.  For example, both the CO-detected
and nondetected target galaxies fall along a common locus in a
500\um/350\um--350\um/250\um\ color-color diagram, but the galaxies
are well mixed in redshift along that locus.

% fig:ratioDetZ
A more general method of estimating redshifts is to take SEDs of
galaxies with known redshift and temperature as templates, then find
which redshifted template best matches the observed SED.  Template
fits are not very tightly constrained by SPIRE data alone because the
250, 350, and 500\um\ bands lie near the peak of the observed-frame
SED, however.  Flux ratios near the peak have little dynamic range and
consequently cannot provide strong constraints: the maximum $S_\nu
(350)/S_\nu (500)$ and $S_\nu (350)/S_\nu (250)$ ratios are about 1.6.

\begin{figure} 
\centering \includegraphics[width=3.2in, clip=true]{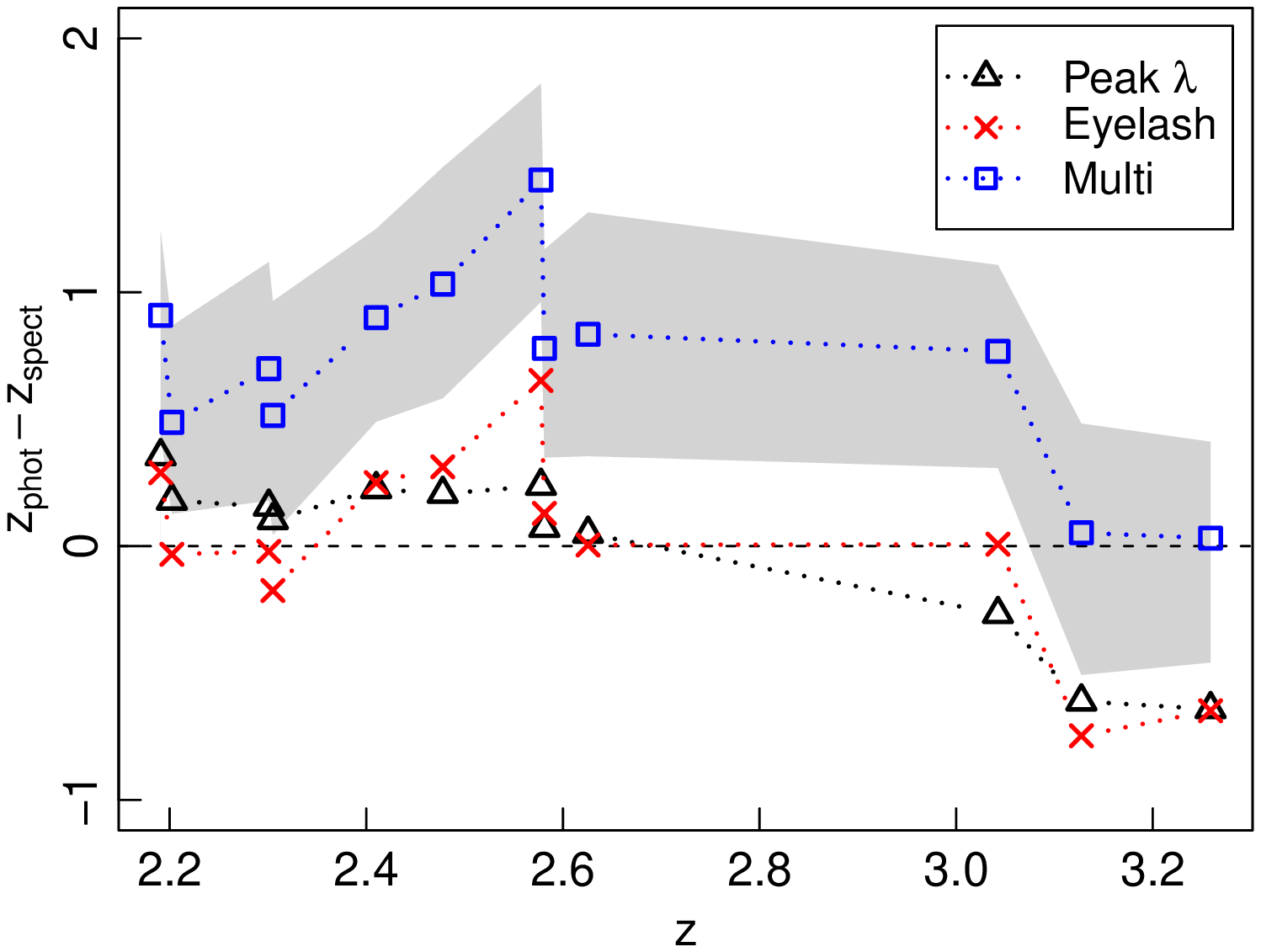}
\caption{Photometric redshift errors versus spectroscopic redshift.
  For the sources with CO $J = 1-0$ redshifts, either the linear fit
  (triangles; Fig.~\ref{fig:ratioDetZ}) or template fitting against the
  Eyelash SED (crosses) provides redshift estimates within $\Delta z
  \approx 0.3$ for $z \leq 3.1$.  Dashed lines highlight grouping by
  method only.  Template fits that automatically
  select from a library of SEDs (squares) generally predict
  redshift to $\Delta z \approx 0.5$-1; the grey background shows the typical internal
  uncertainty for the template fits.  Models that choose between
  multiple SEDs predict redshifts
  outside the Zpectrometer band for several of the nondetected
  sources, even though the linear and Eyelash templates place them 
  within the band, and have lower errors for the $z > 3$
  sources.  As discussed in the text, this indicates that selection
  effects are important in the
  agreement between Eyelash SED and CO detections.
  \label{fig:templCmp}}
\end{figure}

Figure~\ref{fig:templCmp} shows the redshift error $\Delta z$ versus
redshift $z$ for the detected galaxies comparing results from the
simple $S_\nu (250)/S_\nu (500)$ ratio and those from independent
template fitting codes by co-authors Clements, Gonz\'{a}lez-Nuevo, and
Wardlow.  Each estimates the redshift by minimizing the chi-squared
deviation of the data points compared with one or more observed SEDs
of galaxies with known redshifts.  Minimizing chi-squared by shifting
the templates in wavelength gives the redshift, while the widths of
the chi-squared values versus shift provide redshift error estimates.
Comparing the template fitting with the linear fit between $z$ and
$S_\nu (250)/S_\nu (500)$ shown in Fig.~\ref{fig:ratioDetZ}, shows
that the linear fit gives the lowest dispersion in $\Delta z$ for most
of the detected galaxies.  Redshifts from an SED corresponding to the
Cosmic Eyelash galaxy SMM\,J2135$-$0102
\citetext{\citealp{gonzaleznuevo12}, with SED template from
  \citealp{ivison10} and \citealp{swinbank10}} are nearly as accurate
as those from the linear fit.  Comparison with models that choose from
libraries of extragalactic SEDs shows these have generally poorer
agreement than constrained models for most of the detected galaxies,
with $\Delta z \approx 0.5$--1 and internal error estimates smaller
than the actual redshift deviation between observation and model.
Their errors are lower than those of the $S_\nu (250)/S_\nu (500)$
ratio or Eyelash fits for the two galaxies at $z>3.1$, however.  This
result emphasizes the difficulty of fitting curves with only a few
samples near the peak; photometric redshifts incorporating
substantially longer-wavelength data are substantially more accurate
\citep[see, e.g., photometric redshifts including MAMBO 1200\um\ data
in][]{negrello10}.

We emphasize that the correspondence between CO $J = 1-0$ detections
and SPIRE 250\um/500\um\ color or Eyelash template redshifts yields no
redshift information about the galaxies that we do {\em not} detect in
the CO $J = 1-0$ line.  Galaxies with and without detections span
nearly the same 250\um/500\um\ flux density ratios
(Fig.~\ref{fig:ratioDetZ}), and the success of the Eyelash template
may simply arise from its ability to identify the $T \sim 35$\,K
galaxies we find in CO.  As we discussed in
section~\ref{sec:completeness}, some fraction of galaxies with
continuum properties similar to galaxies with CO detections very
likely fall outside the Zpectrometer's band.

\subsection{Lens magnifications}
Lens magnifications ($\mu$) are needed to convert the observed fluxes
to intrinsic luminosities.  Determining magnification is generally
done by constraining a model of the source-lens pair with an image of
the gravitationally-distorted source galaxy.  Detailed lens models are
not yet available for most of our sources.  We do have
velocity-resolved spectra, however, and can take an alternative
approach of comparing observed and intrinsic luminosities established
by an empirical luminosity-linewidth relationship similar in spirit to
the Tully-Fisher relation \citep{tully77}.  Without correction for
galaxy inclination or dispersion in intrinsic galaxy properties, such
a relationship cannot be exact, but it can provide approximate
estimates of lensing magnifications.

Since the galaxies with Zpectrometer CO detections seem to be quite
typical of the general SMG population, as we discussed above, we
assume that only lens magnifications modify the observed typical
luminosity distributions of galaxies detected in CO $J = 1-0$.
Following the method of \citet{bothwell11}, as outlined below, we find
an empirical intrinsic integrated line luminosity-linewidth
relationship $L^\prime = a \left( \Delta v_{\rm FWHM} \right)^b$ from
galaxies with published CO $J = 1-0$ intensities, widths, and
magnifications.  This relationship forms the basis to solve for the
unknown magnification $\mu$ that scales the true line luminosity
$L^\prime$ to an apparent luminosity $L^\prime_{\rm apparent} = \mu
L^\prime$, or
\begin{equation}
   \mu = \frac{L^\prime_{\rm apparent}}{L^\prime} 
   = \frac{L^\prime_{\rm apparent}}{a \left( \Delta v_{\rm FWHM} \right)^b} \;.
\label{eq:mu1}
\end{equation}
 
\begin{figure} 
%\epsscale{0.6}
%\plotone{figs/CO.TF.eps} %fig7, CO.TF.eps
\centering \includegraphics[width=3.2in, clip=true]{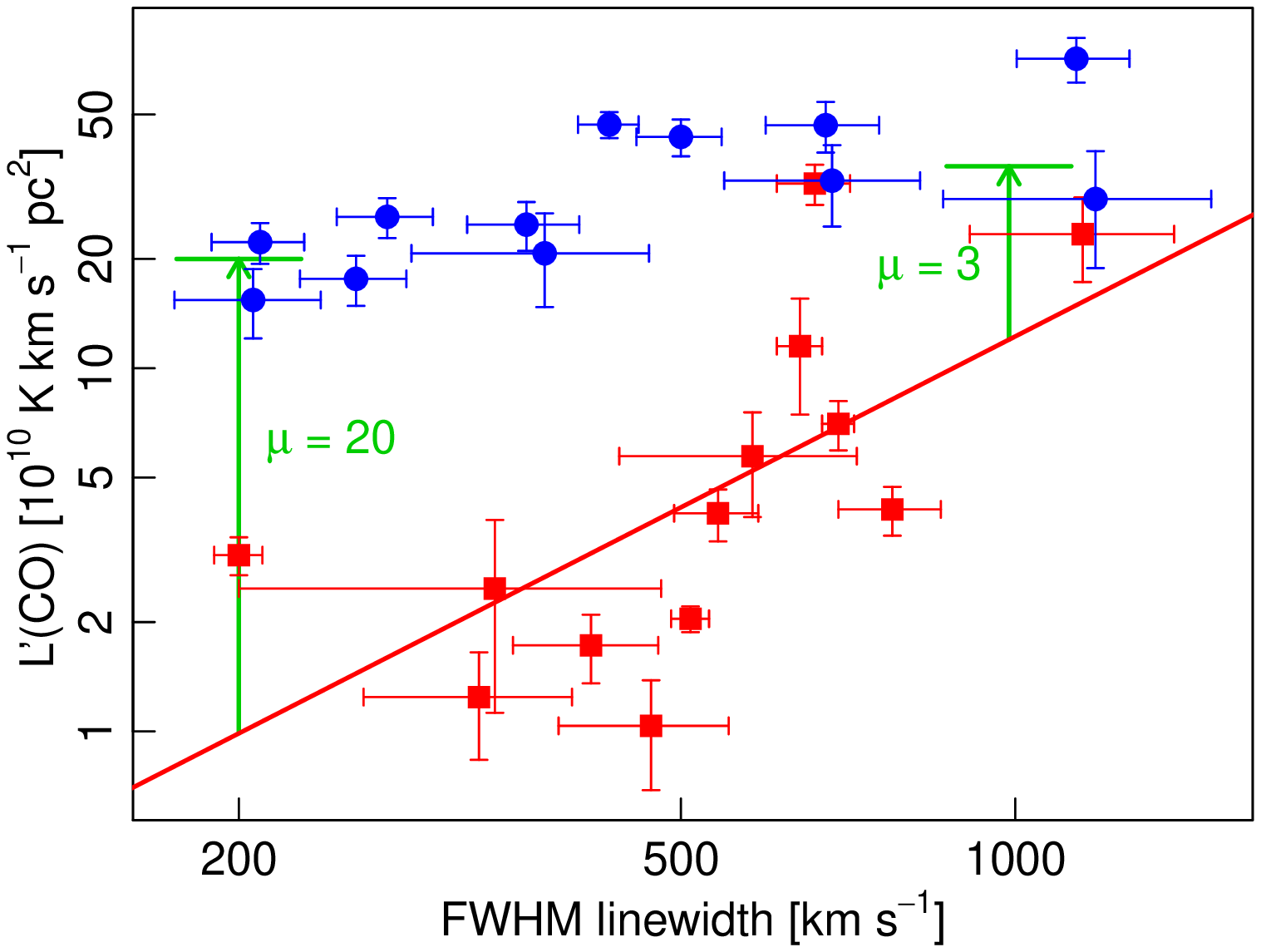}
\caption{Square points indicate line luminosities 
  $L^\prime_{\rm CO}$ versus CO $J = 1-0$ {\rm FWHM} linewidths
  for 15 sources with redshifts, fluxes, and  magnifications as needed
  from the literature \citetext{CO data from
  \citealp{harris10, carilli10, ivison11, riechers11c}; magnifications
  as needed from \citealp{smail02}}.
    The solid line shows a power-law fit to these data, representing an
  estimate of the intrinsic line luminosity versus linewidth for
  typical galaxies.
  Round points show the observed line luminosities versus widths for 
  the Zpectrometer sample (values from
  Table~\ref{tab:summary}), without correction for lens magnifications.  
  The
  literature sample has a steep power-law relation with modest scatter,
  while the Zpectrometer sample
  shows a much flatter trend.  Intrinsically weaker
  sources require  
  higher magnification to rise above the observational detection
  threshold, which is approximately constant in flux.
  We scaled the observed line luminosity for each of the Zpectrometer 
  sources by the
  prediction from the power-law estimate of
  its likely magnification $\mu$, as given in
  Table~\ref{tab:galProps}.  
  This method
  yielded a range of $\mu \approx 3$--20 
  and a median of $\mu = 10$ for the Zpectrometer sample.
  \label{fig:co_tf}}
\end{figure}

Figure~\ref{fig:co_tf} shows $L^\prime _{{\rm CO} J=1-0}$ for 15 $z =
2$--4 SMGs from the literature versus linewidth, with corrections for
lens magnification when needed \citetext{CO data from
  \citealp{harris10, carilli10, ivison11, riechers11c}; magnifications
  as needed from \citealp{smail02}}.  Points for these galaxies fall
close to a power law fit with relatively small scatter.  Inserting the
fit parameters into equation~(\ref{eq:mu1}), we obtain
\begin{equation}
\mu = 3.5 \, \frac{L^\prime_{\rm apparent}}{10^{11} \, {\rm K \, km
      \, s}^{-1} {\rm \, pc}^2} 
    \left( \frac{400 \, {\rm km \,s}^{-1}}{\Delta v_{\rm FWHM}}
    \right)^{1.7} \; ,
\label{eq:mu}
\end{equation}
with the units for line luminosity $L^\prime$ as defined in
\citet{sdr92}.  Apparent line luminosities for the H-ATLAS galaxies
with Zpectrometer CO measurements, uncorrected for lens magnification,
are also plotted in Figure~\ref{fig:co_tf} .  As expected for a
flux-limited sample, these galaxies have a narrower range of
luminosities than the comparison sample from the literature.  Galaxies
with narrower linewidths, which will generally have lower
luminosities, require more magnification to reach the observational
detection threshold.

Table~\ref{tab:galProps} lists the lens magnifications derived from
equation~(\ref{eq:mu}).  The empirical magnification distribution is
approximately uniform over its range of 3 to 20, with both median and
mean equal to 10.  Simple uncertainty estimates for the scaling term
and power law index are not available because the two are not
independent, but a bootstrap analysis provides $\sim$68\% confidence
bounds on $\mu$ for each galaxy in Table~\ref{tab:galProps}.
Fractional uncertainties range from about 30\% at a minimum near
500\kms, climbing to about 50\% at 300 and 1200\kms, and about 100\%
at 200\kms.  The divergence toward low $\Delta v_{\rm FWHM}$ is also a
reminder that the (unknown) inclination corrections become large for
galaxies with observed low linewidths.  It is possible that some of
the sources with broad CO lines and low inferred amplifications from
equation (\ref{eq:mu}) are intrinsically hyper-luminous systems which
are not significantly magnified at all; the magnification uncertainty
encompasses that possibility.

\begin{deluxetable*}{lrrrrrrrr}
\tabletypesize{\scriptsize}
%\rotate
\tablecaption{List of galaxies with observed CO line luminosities and
  estimates for lens magnifications $\mu$ and corrected (intrinsic)
  dust luminosities and gas masses.  The median value of $\mu = 10$.
  Mass calculations are from $M = \alpha \, L^\prime$, with $\alpha =
  0.8$~$M_\odot$~(K~km~s$^{-1}$~pc$^2$)$^{-1}$ \citep{downes98} to
  ease comparisons using this value as a standard.  Masses from $J
  = 1-0$ intensities are likely to provide larger masses by a factor
  between 1.5 and 2 compared with those derived from millimeter-wave
  observations \citep[discussion and references in][]{harris10}.
  Errors do not include uncertainties in $\alpha$ \citep[see,
  e.g.,][for a review]{tacconi08}.  For comparison with other work
  that assumes $\mu \approx 10$, the last two columns contain
  luminosity and mass estimates with $\mu = 10$, with errors
  incorporating a nominal $\Delta\mu/\mu = 0.5$. \label{tab:galProps}}

\tablewidth{0pt} 
\tablecolumns{9}
\tablehead{ \colhead{H-ATLAS Name} &
%\colhead{$\Delta v_{\rm FWHM}$} &
\colhead{$L^\prime_{\rm obs, CO} \, [10^{10}$ K} &
\colhead{$\mu$} &
\colhead{$L_{IR}/\mu$} &
\colhead{$M_{\rm gas}/\mu$} &
\colhead{$L_{IR}/10$} &
\colhead{$M_{\rm gas}/10$} &
\colhead{Target}\\
\colhead{} &
%\colhead{$[$km~s$^{-1}]$} &
\colhead{km s$^{-1}$ pc$^2]$} &
\colhead{} &
\colhead{$[10^{12} \; L_\odot$]} &
\colhead{$[10^{10} \; M_\odot$]} &
\colhead{$[10^{12} \; L_\odot$]} &
\colhead{$[10^{10} \; M_\odot$]} &
\colhead{No.}
}
\startdata
J084933.4+021443 & $29.2 \pm 10.4$ & $2 \pm 1$ & $28.8 \pm 19.6$ & $11.7 \pm 8.0$ & $5.8 \pm 3.5$ & $2.3 \pm 1.4$ & $2$ \\ 
J090302.9$-$014127 & $26.1 \pm 3.3$ & $18 \pm 8$ & $3.8 \pm 1.9$ & $1.2 \pm 0.6$ & $6.9 \pm 3.5$ & $2.1 \pm 1.1$ & $a$ \\ 
J090311.6+003906 & $46.9 \pm 3.8$ & $14 \pm 4$ & $4.1 \pm 1.2$ & $2.7 \pm 0.8$ & $5.8 \pm 2.9$ & $3.8 \pm 1.9$ & $c$ \\ 
J091305.0$-$005343 & $24.9 \pm 3.8$ & $10 \pm 4$ & $3.1 \pm 1.2$ & $2.0 \pm 0.8$ & $3.1 \pm 1.6$ & $2.0 \pm 1.0$ & $b$ \\ 
J091840.8+023047 & $32.9 \pm 8.3$ & $5 \pm 1$ & $8.0 \pm 2.9$ & $5.3 \pm 1.9$ & $4.0 \pm 2.2$ & $2.6 \pm 1.5$ & $5$ \\ 
J113243.1$-$005108 & $20.7 \pm 6.0$ & $8 \pm 3$ & $3.0 \pm 1.3$ & $2.1 \pm 0.9$ & $2.4 \pm 1.4$ & $1.7 \pm 1.0$ & $10$ \\ 
J113526.3$-$014605 & $15.4 \pm 3.3$ & $17 \pm 11$ & $7.1 \pm 4.8$ & $0.7 \pm 0.5$ & $12.1 \pm 6.6$ & $1.2 \pm 0.7$ & $9$ \\ 
J114637.9$-$001132 & $46.7 \pm 7.4$ & $7 \pm 2$ & $22.4 \pm 6.9$ & $5.3 \pm 1.6$ & $15.7 \pm 8.2$ & $3.7 \pm 2.0$ & $13$ \\ 
J115820.2$-$013753 & $17.6 \pm 2.8$ & $13 \pm 7$ & $1.9 \pm 1.0$ & $1.1 \pm 0.6$ & $2.5 \pm 1.3$ & $1.4 \pm 0.7$ & $15$ \\ 
J133649.9+291801 & $22.2 \pm 2.9$ & $23 \pm 15$ & $2.5 \pm 1.6$ & $0.8 \pm 0.5$ & $5.7 \pm 2.9$ & $1.8 \pm 0.9$ & $20$ \\ 
J134429.4+303036 & $71.2 \pm 10.0$ & $4 \pm 2$ & $25.2 \pm 14.2$ & $14.2 \pm 8.0$ & $10.1 \pm 5.2$ & $5.7 \pm 3.0$ & $19$ \\ 
J141351.9$-$000026 & $43.4 \pm 5.0$ & $10 \pm 3$ & $4.9 \pm 1.3$ & $3.5 \pm 0.9$ & $4.9 \pm 2.5$ & $3.5 \pm 1.8$ & $21$
\enddata
\end{deluxetable*}

Magnifications from the luminosity-linewidth relation (\ref{eq:mu})
compare well with values from the two galaxies in the H-ATLAS survey
that have both Zpectrometer detections and preliminary lens models
from image reconstruction with lens and source galaxies at known
redshifts.  For J090311.6+003906 \citep[ID.81; CO data
from][]{frayer11}, the CO linewidth-luminosity relationship in
equation (\ref{eq:mu}) predicts $\mu = 14 \pm 4$, while
\citet{negrello10} quote $\mu \sim 19$ (range 18--31) from their
modeling.  For J091305.0$-$005343 (ID.130), equation (\ref{eq:mu})
predicts $\mu = 10 \pm 4$ while \citet{negrello10} quote $\mu \sim 6$
(range 5--7).

Table~\ref{tab:galProps} provides estimates of the intrinsic
luminosities and gas masses corrected by the source-specific
magnification factors, and also by the median $\mu = 10$.  Most of the
galaxies appear to have intrinsic luminosities typical of ULIRGs,
although three are candidate hyper-LIRGs, with luminosities greater
than $10^{13}\,L_\odot$.  Gas mass estimates use the same CO
luminosity-to-mass conversion factor as in \citet{greve05} and
\citet{tacconi06} for ease of comparison, $\alpha = 0.8$
M$_\odot$~(K\,\kms\,pc$^2$)$^{-1}$ \citep{downes98}.  Although
absolute mass estimates carry double uncertainties from both
conversion and magnification factors, they are similar to estimates
for SMGs selected from 850\um\ observations that use the same conversion
factor.

%tab:galProps

\section{Discussion \& Conclusions}\label{sec:disc}
Demonstrating an ability to rapidly establish spectroscopic redshifts
and molecular line parameters for luminous high-redshift galaxies is a
major step forward in exploring the properties of DSFGs.  In a single
observing season we have added a dozen galaxies with CO $J = 1-0$ line
detections to the pool of sources for study among the bright ``350\um\
peakers'' from the {\it Herschel}-ATLAS survey.  

We established accurate redshifts for sufficient galaxies to compare
with an independent survey of SMGs with observed optical redshifts.
That survey started with submillimeter source continuum detections in
$\sim 15$ arcsecond beams at 850\um, associated those with 20\,cm
radio continuum sources to obtain positions to $\sim 1$ arcsecond, and
then used optical spectroscopy to determine redshifts \citep[see
e.g.,][]{chapman05}.  Subsequent observations with a millimeter-wave
interferometer used the optical redshifts as starting points to
determine molecular redshifts from the star-forming gas itself
\citep[e.g.,][]{neri03, greve05, tacconi06}, conclusively connecting
the 850\um\ sources with massive molecular clouds in the target
galaxies in about half of the attempts.  The Zpectrometer CO
redshifts, in contrast, started with bright 350\um\ sources with
submillimeter photometry indicating rough photometric redshifts, but
then made a single step from continuum to molecular line observations
with this wideband spectrometer.  We find nearly identical redshift
distributions for the two surveys, indicating that each sample is
representative of a general DSFG population.  Coupled with mm-wave
molecular spectroscopy of many of the brighter 850\um\ sources
\citep[e.g.,][]{neri03, greve05, tacconi06}, the agreement in
distributions between independent samples lays to rest some of the
concern that SMG redshift distributions from optical spectroscopy may
be influenced by misidentifications between the submillimeter sources
and counterpart optical galaxies.

Dust temperatures and linewidths of the galaxies with CO 1--0 detections
appear to be similar to SMGs.  With spectroscopic redshifts, we are
able to break the temperature-redshift degeneracy to find a
characteristic temperature of 34\,K (in an optically thin formalism),
similar to those typically reported for SMGs \citep{kovacs06,
  chapman10, wardlow11}.  Galaxies with redshifts $z > 3$ may be
somewhat warmer than those at lower redshift.  The CO linewidths have
very similar distributions for these H-ATLAS CO $J = 1-0$ detections
and for SMGs observed in mid-$J$ lines.  Both of these comparisons add
further to an argument that the bulk of samples selected by bright
850\um\ and SPIRE emission draw from the same parent population.

\pagebreak
Most, if not all, of the galaxies we observed appear to be
gravitationally lensed.  We estimate lens magnifications with an
empirical CO linewidth-luminosity relationship made possible with
velocity-resolved spectroscopy.  We verified magnification agreement
within factors close to two for two galaxies with published lens
modeling.  We find magnifications from 3 to 20, a range matching those
for lenses with detailed models, $\mu \approx 5$--30
\citep{swinbank10, negrello10, gavazzi11, lestrade11}.  In the absence
of other information, a typical value of $\mu = 10$ is likely to be
correct within a factor of two for the bright H-ATLAS galaxies.

With even moderately accurate lens magnification estimates, we can
make reasonable estimates of intrinsic luminosities and gas masses of
the galaxies in our sample (Table~\ref{tab:galProps}).  Most have
luminosities characteristic of ULIRGs, but there are three targets
that could be hyper-LIRGs with luminosities $> 10^{13}\,L_\odot$.
The most luminous of these could well be multiple galaxies within 
the beam, as their linewidths are broad.  Even accounting for lensing
magnification, however, the bright H-ATLAS galaxies seem
to be drawn from the high-luminosity, high-mass end of the DSFG
distribution.

The overall agreement in properties between the Zpectrometer and
850--1200\um-selected SMG samples indicates that the two populations
are drawn from the same distribution.  In principle, the additional
selection introduced by lens magnification could have emphasized
different properties from the SMG population, but it does not appear
to have done so.  Lensed sources will be valuable in understanding the
properties of the underlying DSFG population, including the SMGs: for
observations, magnification by even a factor of a few translates to an
order of magnitude reduction in integration time.  Detailed studies of
the galaxies' interstellar media in multiple molecular and atomic
lines becomes feasible in the bright lensed sample
\citep[e.g.,][]{danielson11}, with magnification also increasing
spatial resolution in the source plane \citep[e.g.,][]{swinbank10}.

\pagebreak
Overall, the detection fraction of about half of the targets compares
well with other programs that observed molecular lines from high
redshift galaxies.  A conservative extrapolation of a CO-to-continuum
intensity relationship indicates that the nondetection rate is
substantially higher than expected from chance alone, evidence that
some of the simple photometric redshifts are incorrect.  Our selection
is from SPIRE photometry identification of ``350\um\ peakers.''
Comparison of the SPIRE data and CO redshifts indicates that the
continuum peak wavelength, deduced from the 250\um/500\um\ flux
density ratio, predicts the redshift of the galaxies from our
carefully-selected sample with $\Delta z \sim 0.3$.  The SPIRE colors
do not necessarily provide accurate estimates for galaxies we do not
detect, however.  A simple explanation for some of the nondetections
is emission from dust warmer than the $\sim 35$\,K characteristic for
most of the galaxies with CO detections.  The redshift-temperature
degeneracy for the observed peak emission wavelength, $\lambda_{\rm
  peak, obs} \propto T/(1+z)$ \citep[e.g.,][]{blain99degen} causes
problems for photometric selection in the SPIRE bands alone: warmer
galaxies with redshift above the Zpectrometer's upper limit will
appear in the target list.  It is very likely that this is the reason
for the nondetection of the target with third-brightest 350\um\ flux
density in this study.

Reasons other than warm galaxies or CO intensities that fall below our
observational limit can also contribute to nondetections. Emission
from multiple galaxies within the beam at one or more wavelengths can
contribute flux to distort the apparent SED, leading to erroneous
redshifts from single-component models.  This is likely to affect
weaker sources preferentially, both because smaller absolute amounts
of contaminating flux are increasingly important and because the
density of contaminating sources increases with decreasing flux.
Multiple distinct temperature components within a source, for instance
a central AGN-heated dust region surrounded by less excited material,
will also distort the SED.  Depending upon the temperature structure
within the source and the size and position of the source relative to
any lens caustic, differential amplification could distort observed
SEDs.  SEDs in all these cases will be poorly represented by the
simple single-component model we have used in target selection, and
photometric redshifts based on SPIRE data alone will fail to some
extent.

The observed redshift distribution and the dust SED properties hint at
some change in DSFG properties near $z \sim 3$.  The redshift
cumulative distribution functions for both the Zpectrometer and
\citet{chapman05} samples show a slope break at $z \sim 2.9$ although
they are selected in different ways.  This indicates that DSFGs become
less common, or are selected less efficiently by continuum properties,
above that redshift.  An increase in the number of warm galaxies with
redshift would both explain the tentative trend in temperature we see
and explain the lower CO detection rate.  Temperature alone may not
suffice to explain the break, and explanations involving more complex
SEDs with increasing redshift are attractive as well.
\citet{wardlow11} use extensive optical and infrared photometry on a
sample of 74 870\um-selected SMGs, also finding a precipitous drop in
submillimeter source counts above $z_{\rm phot} \sim 2.8$ in spite of
their sensitivity to emission from warm dust.  Increased sample sizes
are necessary to establish whether there is a real change in DSFG
properties near $z \sim 3$ or not.

The CO data presented here are among the earliest spectroscopic
explorations of {\em Herschel}-selected galaxies.  Continuing
observations of the CO $J = 1-0$ line by the GBT and Jansky-VLA will
provide a more diverse sample of galaxies: some from broadened
continuum selection criteria, and some with deeper integrations to
reach weaker line emission.  Wideband spectrometers now coming on line
for the 3\,mm band at the LMT and IRAM 30\,m will also add to the
number of galaxies with known molecular redshifts.  An expanded sample
will strengthen our understanding of the relationship between
molecular and continuum emission in DSFGs.

%%%%%%%%%%%%%%%%%

\acknowledgments We thank L.~Leeuw, M.~Michalowski, and I.~Valtchanov
for their comments on aspects of this work.  We acknowledge support
from the National Science Foundation under grant numbers AST-0503946
to the University of Maryland, AST-0708653 to Rutgers University.  DTF
acknowledges support by NASA through an award issued by JPL/Caltech;
DR acknowledges support from NASA through a Spitzer Space Telescope
Grant; IRS and AMS acknowledge support from STFC; SB acknowledges
financial contribution from the agreement ASI-INAF I/009/10/0; JGN
acknowledges financial support from Spanish CSIC for a JAE-DOC
fellowship and partial financial support from the Spanish Ministerio
de Ciencia e Innovacion project AYA2010-21766-C03-01.  Results here
came from GBT programs 8C-09 (PI Smail), 9A-40 (PI Swinbank), 10C-29
(PI Frayer), and 11A-27 (PI Frayer).  We thank the GBT staff for their
support and contributions.  The National Radio Astronomy Observatory
is a facility of the National Science Foundation operated under
cooperative agreement by Associated Universities, Inc.  The {\em
  Herschel}-ATLAS is a project with {\em Herschel}, which is an ESA
space observatory with science instruments provided by European-led
Principal Investigator consortia and with important participation from
NASA. The H-ATLAS website is http://www.h-atlas.org/.

{\em Facilities:} \facility{GBT}, \facility{{\em Herschel Space Observatory}}.


\begin{thebibliography}{98}
\expandafter\ifx\csname natexlab\endcsname\relax\def\natexlab#1{#1}\fi

\bibitem[{{Alexander} {et~al.}(2005){Alexander}, {Bauer}, {Chapman}, {Smail},
  {Blain}, {Brandt}, \& {Ivison}}]{alexander05}
{Alexander}, D.~M., {Bauer}, F.~E., {Chapman}, S.~C., {Smail}, I., {Blain},
  A.~W., {Brandt}, W.~N., \& {Ivison}, R.~J. 2005, \apj, 632, 736

\bibitem[{{Alexander} {et~al.}(2003)}]{alexander03}
{Alexander}, D.~M., {et~al.} 2003, \aj, 125, 383

\bibitem[{Amblard {et~al.}(2010)}]{amblard10}
Amblard, A., {et~al.} 2010, \aap, 518, L9

\bibitem[{{Austermann} {et~al.}(2010)}]{austermann10}
{Austermann}, J.~E., {et~al.} 2010, \mnras, 401, 160

\bibitem[{{Baker} {et~al.}(2007){Baker}, {Stacey}, {Glenn}, {Erickson},
  {Harris}, \& {Wootten}}]{zmachines07}
{Baker}, A., {Stacey}, G., {Glenn}, J., {Erickson}, N., {Harris}, A., \&
  {Wootten}, A. 2007, in Astronomical Society of the Pacific Conference Series,
  Vol. 375, From Z-Machines to ALMA: (Sub)Millimeter Spectroscopy of Galaxies,
  ed. {A.~J.~Baker, J.~Glenn, A.~I.~Harris, J.~G.~Mangum, \& M.~S.~Yun }, 3

\bibitem[{{Barger} {et~al.}(1998){Barger}, {Cowie}, {Sanders}, {Fulton},
  {Taniguchi}, {Sato}, {Kawara}, \& {Okuda}}]{barger98}
{Barger}, A.~J., {Cowie}, L.~L., {Sanders}, D.~B., {Fulton}, E., {Taniguchi},
  Y., {Sato}, Y., {Kawara}, K., \& {Okuda}, H. 1998, \nat, 394, 248

\bibitem[{{Baugh} {et~al.}(2005){Baugh}, {Lacey}, {Frenk}, {Granato}, {Silva},
  {Bressan}, {Benson}, \& {Cole}}]{baugh05}
{Baugh}, C.~M., {Lacey}, C.~G., {Frenk}, C.~S., {Granato}, G.~L., {Silva}, L.,
  {Bressan}, A., {Benson}, A.~J., \& {Cole}, S. 2005, \mnras, 356, 1191

\bibitem[{{Blain}(1999)}]{blain99degen}
{Blain}, A.~W. 1999, \mnras, 309, 955

\bibitem[{{Blain} {et~al.}(2003){Blain}, {Barnard}, \& {Chapman}}]{blain03}
{Blain}, A.~W., {Barnard}, V.~E., \& {Chapman}, S.~C. 2003, \mnras, 338, 733

\bibitem[{{Blain} {et~al.}(2004){Blain}, {Chapman}, {Smail}, \&
  {Ivison}}]{blain04}
{Blain}, A.~W., {Chapman}, S.~C., {Smail}, I., \& {Ivison}, R. 2004, \apj, 611,
  52

\bibitem[{{Blain} {et~al.}(2002){Blain}, {Smail}, {Ivison}, {Kneib}, \&
  {Frayer}}]{blain02}
{Blain}, A.~W., {Smail}, I., {Ivison}, R.~J., {Kneib}, J., \& {Frayer}, D.~T.
  2002, \physrep, 369, 111

\bibitem[{{Bothwell} {et~al.}(2012)}]{bothwell11}
{Bothwell}, M.~S., {et~al.} 2012, in prep.

\bibitem[{{Capak} {et~al.}(2008)}]{capak08}
{Capak}, P., {et~al.} 2008, \apjl, 681, L53

\bibitem[{{Capak} {et~al.}(2011)}]{capak11}
{Capak}, P.~L., {et~al.} 2011, \nat, 470, 233

\bibitem[{{Carilli} {et~al.}(2010)}]{carilli10}
{Carilli}, C.~L., {et~al.} 2010, \apj, 714, 1407

\bibitem[{{Chapman} {et~al.}(2003){Chapman}, {Blain}, {Ivison}, \&
  {Smail}}]{chapman03}
{Chapman}, S.~C., {Blain}, A.~W., {Ivison}, R.~J., \& {Smail}, I.~R. 2003,
  \nat, 422, 695

\bibitem[{{Chapman} {et~al.}(2005){Chapman}, {Blain}, {Smail}, \&
  {Ivison}}]{chapman05}
{Chapman}, S.~C., {Blain}, A.~W., {Smail}, I., \& {Ivison}, R.~J. 2005, \apj,
  622, 772

\bibitem[{{Chapman} {et~al.}(2004){Chapman}, {Smail}, {Blain}, \&
  {Ivison}}]{chapman04}
{Chapman}, S.~C., {Smail}, I., {Blain}, A.~W., \& {Ivison}, R.~J. 2004, \apj,
  614, 671

\bibitem[{{Chapman} {et~al.}(2010)}]{chapman10}
{Chapman}, S.~C., {et~al.} 2010, \mnras, 409, L13

\bibitem[{{Conselice} {et~al.}(2003){Conselice}, {Chapman}, \&
  {Windhorst}}]{conselice03}
{Conselice}, C.~J., {Chapman}, S.~C., \& {Windhorst}, R.~A. 2003, \apjl, 596,
  L5

\bibitem[{{Coppin} {et~al.}(2010{\natexlab{a}})}]{coppin10midir}
{Coppin}, K., {et~al.} 2010{\natexlab{a}}, \apj, 713, 503

\bibitem[{{Coppin} {et~al.}(2009)}]{coppin09}
{Coppin}, K.~E.~K., {et~al.} 2009, \mnras, 395, 1905

\bibitem[{{Coppin} {et~al.}(2010{\natexlab{b}})}]{coppin10co}
---. 2010{\natexlab{b}}, \mnras, 407, L103

\bibitem[{{Daddi} {et~al.}(2010)}]{daddi10}
{Daddi}, E., {et~al.} 2010, \apjl, 714, L118

\bibitem[{{Danielson} {et~al.}(2011)}]{danielson11}
{Danielson}, A.~L.~R., {et~al.} 2011, \mnras, 410, 1687

\bibitem[{{Dannerbauer} {et~al.}(2004){Dannerbauer}, {Lehnert}, {Lutz},
  {Tacconi}, {Bertoldi}, {Carilli}, {Genzel}, \& {Menten}}]{dannerbauer04}
{Dannerbauer}, H., {Lehnert}, M.~D., {Lutz}, D., {Tacconi}, L., {Bertoldi}, F.,
  {Carilli}, C., {Genzel}, R., \& {Menten}, K.~M. 2004, \apj, 606, 664

\bibitem[{{Dav{\'e}} {et~al.}(2010){Dav{\'e}}, {Finlator}, {Oppenheimer},
  {Fardal}, {Katz}, {Kere{\v s}}, \& {Weinberg}}]{dave10}
{Dav{\'e}}, R., {Finlator}, K., {Oppenheimer}, B.~D., {Fardal}, M., {Katz}, N.,
  {Kere{\v s}}, D., \& {Weinberg}, D.~H. 2010, \mnras, 404, 1355

\bibitem[{{Downes} \& {Solomon}(1998)}]{downes98}
{Downes}, D., \& {Solomon}, P.~M. 1998, \apj, 507, 615

\bibitem[{{Driver} {et~al.}(2009)}]{driver09}
{Driver}, S.~P., {et~al.} 2009, in IAU Symposium, Vol. 254, IAU Symposium, ed.
  {J.~Andersen, J.~Bland-Hawthorn, \& B.~Nordstr{\"o}m}, 469--474

\bibitem[{Eales {et~al.}(2010)}]{eales10}
Eales, S., {et~al.} 2010, \pasp, 122, 499

\bibitem[{{Fixsen} {et~al.}(1998){Fixsen}, {Dwek}, {Mather}, {Bennett}, \&
  {Shafer}}]{fixsen98}
{Fixsen}, D.~J., {Dwek}, E., {Mather}, J.~C., {Bennett}, C.~L., \& {Shafer},
  R.~A. 1998, \apj, 508, 123

\bibitem[{{Frayer} {et~al.}(1999){Frayer}, {Ivison}, {Scoville}, {Evans},
  {Yun}, {Smail}, {Barger}, {Blain}, \& {Kneib}}]{frayer99}
{Frayer}, D.~T., {Ivison}, R.~J., {Scoville}, N.~Z., {Evans}, A.~S., {Yun},
  M.~S., {Smail}, I., {Barger}, A.~J., {Blain}, A.~W., \& {Kneib}, J.-P. 1999,
  \apjl, 514, L13

\bibitem[{{Frayer} {et~al.}(1998){Frayer}, {Ivison}, {Scoville}, {Yun},
  {Evans}, {Smail}, {Blain}, \& {Kneib}}]{frayer98}
{Frayer}, D.~T., {Ivison}, R.~J., {Scoville}, N.~Z., {Yun}, M., {Evans}, A.~S.,
  {Smail}, I., {Blain}, A.~W., \& {Kneib}, J. 1998, \apjl, 506, L7

\bibitem[{Frayer {et~al.}(2011)}]{frayer11}
Frayer, D.~T., {et~al.} 2011, \apjl, 726, L22

\bibitem[{Gavazzi {et~al.}(2011)}]{gavazzi11}
Gavazzi, R., {et~al.} 2011, \apj, 738, 125

\bibitem[{{Genzel} {et~al.}(2003){Genzel}, {Baker}, {Tacconi}, {Lutz}, {Cox},
  {Guilloteau}, \& {Omont}}]{genzel03}
{Genzel}, R., {Baker}, A.~J., {Tacconi}, L.~J., {Lutz}, D., {Cox}, P.,
  {Guilloteau}, S., \& {Omont}, A. 2003, \apj, 584, 633

\bibitem[{{Genzel} {et~al.}(2010)}]{genzel10}
{Genzel}, R., {et~al.} 2010, \mnras, 407, 2091

\bibitem[{{Gonz\'{a}lez-Nuevo} {et~al.}(2012)}]{gonzaleznuevo12}
{Gonz\'{a}lez-Nuevo}, J., {et~al.} 2012, ArXiv e-prints, 1202.0402

\bibitem[{Greve {et~al.}(2005)}]{greve05}
Greve, T.~R., {et~al.} 2005, \mnras, 359, 1165

\bibitem[{Griffin {et~al.}(2010)}]{griffin10}
Griffin, M.~J., {et~al.} 2010, \aap, 518, L3

\bibitem[{{Hainline} {et~al.}(2011){Hainline}, {Blain}, {Smail}, {Alexander},
  {Armus}, {Chapman}, \& {Ivison}}]{hainline10}
{Hainline}, L.~J., {Blain}, A.~W., {Smail}, I., {Alexander}, D.~M., {Armus},
  L., {Chapman}, S.~C., \& {Ivison}, R.~J. 2011, ArXiv e-prints,
  astro-ph/1104.4119

\bibitem[{{Harris}(2005)}]{harris05}
{Harris}, A.~I. 2005, Rev. Sci. Inst., 76, 4503

\bibitem[{{Harris} {et~al.}(2010){Harris}, {Baker}, {Zonak}, {Sharon},
  {Genzel}, {Rauch}, {Watts}, \& {Creager}}]{harris10}
{Harris}, A.~I., {Baker}, A.~J., {Zonak}, S.~G., {Sharon}, C.~E., {Genzel}, R.,
  {Rauch}, K., {Watts}, G., \& {Creager}, R. 2010, \apj, 723, 1139

\bibitem[{{Harris et al.}(2007)}]{zp07etal}
{Harris et al.}, A.~I. 2007, in Astronomical Society of the Pacific Conference
  Series, Vol. 375, From Z-Machines to ALMA: (Sub)Millimeter Spectroscopy of
  Galaxies, ed. A.~J. {Baker}, J.~{Glenn}, A.~I. {Harris}, J.~G. {Mangum}, \&
  M.~S. {Yun}, 82--92

\bibitem[{{Hatsukade} {et~al.}(2011)}]{hatsukade11}
{Hatsukade}, B., {et~al.} 2011, \mnras, 411, 102

\bibitem[{{Houck} {et~al.}(2005)}]{houck05}
{Houck}, J.~R., {et~al.} 2005, \apjl, 622, L105

\bibitem[{{Hughes} {et~al.}(1998)}]{hughes98}
{Hughes}, D.~H., {et~al.} 1998, \nat, 394, 241

\bibitem[{{Ibar} {et~al.}(2010)}]{ibar10}
{Ibar}, E., {et~al.} 2010, \mnras, 409, 38

\bibitem[{{Ivison} {et~al.}(2011){Ivison}, {Papadopoulos}, {Smail}, {Greve},
  {Thomson}, {Xilouris}, \& {Chapman}}]{ivison11}
{Ivison}, R.~J., {Papadopoulos}, P.~P., {Smail}, I., {Greve}, T.~R., {Thomson},
  A.~P., {Xilouris}, E.~M., \& {Chapman}, S.~C. 2011, \mnras, 412, 1913

\bibitem[{{Ivison} {et~al.}(2000){Ivison}, {Smail}, {Barger}, {Kneib}, {Blain},
  {Owen}, {Kerr}, \& {Cowie}}]{ivison00}
{Ivison}, R.~J., {Smail}, I., {Barger}, A.~J., {Kneib}, J.-P., {Blain}, A.~W.,
  {Owen}, F.~N., {Kerr}, T.~H., \& {Cowie}, L.~L. 2000, \mnras, 315, 209

\bibitem[{{Ivison} {et~al.}(1998){Ivison}, {Smail}, {Le Borgne}, {Blain},
  {Kneib}, {Bezecourt}, {Kerr}, \& {Davies}}]{ivison98}
{Ivison}, R.~J., {Smail}, I., {Le Borgne}, J.-F., {Blain}, A.~W., {Kneib},
  J.-P., {Bezecourt}, J., {Kerr}, T.~H., \& {Davies}, J.~K. 1998, \mnras, 298,
  583

\bibitem[{{Ivison} {et~al.}(2010){Ivison}, {Smail}, {Papadopoulos}, {Wold},
  {Richard}, {Swinbank}, {Kneib}, \& {Owen}}]{ivison10}
{Ivison}, R.~J., {Smail}, I., {Papadopoulos}, P.~P., {Wold}, I., {Richard}, J.,
  {Swinbank}, A.~M., {Kneib}, J., \& {Owen}, F.~N. 2010, \mnras, 261

\bibitem[{{Ivison} {et~al.}(2005)}]{ivison05}
{Ivison}, R.~J., {et~al.} 2005, \mnras, 364, 1025

\bibitem[{{Kneib} {et~al.}(2005){Kneib}, {Neri}, {Smail}, {Blain}, {Sheth},
  {van der Werf}, \& {Knudsen}}]{kneib05}
{Kneib}, J.-P., {Neri}, R., {Smail}, I., {Blain}, A., {Sheth}, K., {van der
  Werf}, P., \& {Knudsen}, K.~K. 2005, \aap, 434, 819

\bibitem[{{Knudsen} {et~al.}(2008){Knudsen}, {van der Werf}, \&
  {Kneib}}]{knudsen08}
{Knudsen}, K.~K., {van der Werf}, P.~P., \& {Kneib}, J.-P. 2008, \mnras, 384,
  1611

\bibitem[{{Kov{\'a}cs} {et~al.}(2006){Kov{\'a}cs}, {Chapman}, {Dowell},
  {Blain}, {Ivison}, {Smail}, \& {Phillips}}]{kovacs06}
{Kov{\'a}cs}, A., {Chapman}, S.~C., {Dowell}, C.~D., {Blain}, A.~W., {Ivison},
  R.~J., {Smail}, I., \& {Phillips}, T.~G. 2006, \apj, 650, 592

\bibitem[{{Lestrade} {et~al.}(2011){Lestrade}, {Carilli}, {Thanjavur}, {Kneib},
  {Riechers}, {Bertoldi}, {Walter}, \& {Omont}}]{lestrade11}
{Lestrade}, J., {Carilli}, C.~L., {Thanjavur}, K., {Kneib}, J.~., {Riechers},
  D.~A., {Bertoldi}, F., {Walter}, F., \& {Omont}, A. 2011, \apjl, 739, L30

\bibitem[{{Lindner} {et~al.}(2011)}]{lindner11}
{Lindner}, R.~R., {et~al.} 2011, \apj, 737, 83

\bibitem[{Lupu {et~al.}(2010)}]{lupu10}
Lupu, R.~E., {et~al.} 2010, ArXiv e-prints: astro-ph/1009.5983

\bibitem[{{Magdis} {et~al.}(2010)}]{magdis10}
{Magdis}, G.~E., {et~al.} 2010, \mnras, 409, 22

\bibitem[{{Marriage} {et~al.}(2011)}]{marriage11}
{Marriage}, T.~A., {et~al.} 2011, \apj, 731, 100

\bibitem[{{Marsden} {et~al.}(2009)}]{marsden09}
{Marsden}, G., {et~al.} 2009, \apj, 707, 1729

\bibitem[{{Men{\'e}ndez-Delmestre} {et~al.}(2007)}]{menendezdelmestre07}
{Men{\'e}ndez-Delmestre}, K., {et~al.} 2007, \apjl, 655, L65

\bibitem[{{Men{\'e}ndez-Delmestre} {et~al.}(2009)}]{menendezdelmestre09}
---. 2009, \apj, 699, 667

\bibitem[{{Negrello} {et~al.}(2007){Negrello}, {Perrotta},
  {Gonz{\'a}lez-Nuevo}, {Silva}, {de Zotti}, {Granato}, {Baccigalupi}, \&
  {Danese}}]{negrello07}
{Negrello}, M., {Perrotta}, F., {Gonz{\'a}lez-Nuevo}, J., {Silva}, L., {de
  Zotti}, G., {Granato}, G.~L., {Baccigalupi}, C., \& {Danese}, L. 2007,
  \mnras, 377, 1557

\bibitem[{Negrello {et~al.}(2010)}]{negrello10}
Negrello, M., {et~al.} 2010, Science, 330, 800

\bibitem[{{Neri} {et~al.}(2003)}]{neri03}
{Neri}, R., {et~al.} 2003, \apjl, 597, L113

\bibitem[{{Pascale} {et~al.}(2011)}]{pascale11}
{Pascale}, E., {et~al.} 2011, \mnras, 415, 911

\bibitem[{Pilbratt {et~al.}(2010)}]{pilbratt10}
Pilbratt, G.~L., {et~al.} 2010, \aap, 518, L1

\bibitem[{{Pope} {et~al.}(2008)}]{pope08}
{Pope}, A., {et~al.} 2008, \apj, 675, 1171

\bibitem[{{Puget} {et~al.}(1996){Puget}, {Abergel}, {Bernard}, {Boulanger},
  {Burton}, {Desert}, \& {Hartmann}}]{puget96}
{Puget}, J.-L., {Abergel}, A., {Bernard}, J.-P., {Boulanger}, F., {Burton},
  W.~B., {Desert}, F.-X., \& {Hartmann}, D. 1996, \aap, 308, L5

\bibitem[{{R Development Core Team}(2006)}]{Rref}
{R Development Core Team}. 2006, {R}: A Language and Environment for
  Statistical Computing, R Foundation for Statistical Computing, Vienna,
  Austria, http://www.R-project.org

\bibitem[{{Riechers} {et~al.}(2011{\natexlab{a}}){Riechers}, {Hodge}, {Walter},
  {Carilli}, \& {Bertoldi}}]{riechers11c}
{Riechers}, D.~A., {Hodge}, J., {Walter}, F., {Carilli}, C.~L., \& {Bertoldi},
  F. 2011{\natexlab{a}}, \apjl, 739, L31

\bibitem[{{Riechers} {et~al.}(2010)}]{riechers10}
{Riechers}, D.~A., {et~al.} 2010, \apjl, 720, L131

\bibitem[{{Riechers} {et~al.}(2011{\natexlab{b}})}]{riechers11b}
---. 2011{\natexlab{b}}, \apjl, 739, L32

\bibitem[{{Riechers} {et~al.}(2011{\natexlab{c}})}]{riechers11a}
---. 2011{\natexlab{c}}, \apjl, 733, L12

\bibitem[{{Rigby} {et~al.}(2011)}]{rigby11}
{Rigby}, E.~E., {et~al.} 2011, \mnras, 955

\bibitem[{{Scott} {et~al.}(2011)}]{scott11}
{Scott}, K.~S., {et~al.} 2011, \apj, 733, 29

\bibitem[{{Smail} {et~al.}(2004){Smail}, {Chapman}, {Blain}, \&
  {Ivison}}]{smail04}
{Smail}, I., {Chapman}, S.~C., {Blain}, A.~W., \& {Ivison}, R.~J. 2004, \apj,
  616, 71

\bibitem[{{Smail} {et~al.}(1997){Smail}, {Ivison}, \& {Blain}}]{smail97}
{Smail}, I., {Ivison}, R.~J., \& {Blain}, A.~W. 1997, \apjl, 490, L5

\bibitem[{{Smail} {et~al.}(2002){Smail}, {Ivison}, {Blain}, \&
  {Kneib}}]{smail02}
{Smail}, I., {Ivison}, R.~J., {Blain}, A.~W., \& {Kneib}, J.-P. 2002, \mnras,
  331, 495

\bibitem[{{Smith} {et~al.}(2011)}]{smith11}
{Smith}, D.~J.~B., {et~al.} 2011, \mnras, 416, 857

\bibitem[{{Solomon} {et~al.}(1992){Solomon}, {Downes}, \& {Radford}}]{sdr92}
{Solomon}, P.~M., {Downes}, D., \& {Radford}, S.~J.~E. 1992, \apjl, 398, L29

\bibitem[{{Somerville} {et~al.}(2011){Somerville}, {Gilmore}, {Primack}, \&
  {Dominguez}}]{somerville11}
{Somerville}, R.~S., {Gilmore}, R.~C., {Primack}, J.~R., \& {Dominguez}, A.
  2011, ArXiv e-prints, astro-ph/1104.0669

\bibitem[{{Spergel} {et~al.}(2007)}]{spergel07}
{Spergel}, D.~N., {et~al.} 2007, \apjs, 170, 377

\bibitem[{{Swinbank} {et~al.}(2008)}]{swinbank08}
{Swinbank}, A.~M., {et~al.} 2008, \mnras, 391, 420

\bibitem[{{Swinbank} {et~al.}(2010)}]{swinbank10}
---. 2010, \nat, 464, 733

\bibitem[{{Symeonidis} {et~al.}(2011){Symeonidis}, {Page}, \&
  {Seymour}}]{symeonidis11}
{Symeonidis}, M., {Page}, M.~J., \& {Seymour}, N. 2011, \mnras, 411, 983

\bibitem[{{Tacconi} {et~al.}(2006)}]{tacconi06}
{Tacconi}, L.~J., {et~al.} 2006, \apj, 640, 228

\bibitem[{{Tacconi} {et~al.}(2008)}]{tacconi08}
---. 2008, \apj, 680, 246

\bibitem[{{The Astronomical Almanac}(2011)}]{AstAlm2011}
{The Astronomical Almanac}. 2011, {U.S.~Naval Observatory and Rutherford
  Appleton Laboratory}

\bibitem[{{Tully} \& {Fisher}(1977)}]{tully77}
{Tully}, R.~B., \& {Fisher}, J.~R. 1977, \aap, 54, 661

\bibitem[{{Valiante} {et~al.}(2007){Valiante}, {Lutz}, {Sturm}, {Genzel},
  {Tacconi}, {Lehnert}, \& {Baker}}]{valiante07}
{Valiante}, E., {Lutz}, D., {Sturm}, E., {Genzel}, R., {Tacconi}, L.~J.,
  {Lehnert}, M.~D., \& {Baker}, A.~J. 2007, \apj, 660, 1060

\bibitem[{{Vieira} {et~al.}(2010)}]{vieira10}
{Vieira}, J.~D., {et~al.} 2010, \apj, 719, 763

\bibitem[{Wardlow {et~al.}(2011)}]{wardlow11}
Wardlow, J.~L., {et~al.} 2011, \mnras, 414, 917

\bibitem[{{Weedman} {et~al.}(2006)}]{weedman06}
{Weedman}, D., {et~al.} 2006, \apj, 653, 101

\bibitem[{{Wei{\ss}} {et~al.}(2009){Wei{\ss}}, {Ivison}, {Downes}, {Walter},
  {Cirasuolo}, \& {Menten}}]{weiss09}
{Wei{\ss}}, A., {Ivison}, R.~J., {Downes}, D., {Walter}, F., {Cirasuolo}, M.,
  \& {Menten}, K.~M. 2009, \apjl, 705, L45

\bibitem[{{Yan} {et~al.}(2007)}]{yan07}
{Yan}, L., {et~al.} 2007, \apj, 658, 778

\end{thebibliography}
\end{document}